\documentclass{article} % For LaTeX2e
\usepackage{iclr2025_conference,times}

\usepackage[T1]{fontenc}
\usepackage[utf8]{inputenc}
\usepackage{textcomp}
\usepackage{amsmath,amsfonts,bm}

\def\eqref#1{equation~\ref{#1}}
\def\1{\bm{1}}

\DeclareMathAlphabet{\mathsfit}{\encodingdefault}{\sfdefault}{m}{sl}
\SetMathAlphabet{\mathsfit}{bold}{\encodingdefault}{\sfdefault}{bx}{n}

\DeclareMathOperator{\sign}{sign}

\usepackage{import}
\usepackage{hyperref}
\usepackage{url}
\usepackage{xargs} % Required for the \newcommandx
\usepackage{booktabs}
\usepackage{titling}
\usepackage{float}
\usepackage{graphicx}
\usepackage{textcomp}
\usepackage{geometry}
\newcommand{\dprime}{^{\prime\prime}}

\DeclareMathOperator{\log1p}{log1p}

\makeatother
\title{Surya: Foundation Model for Heliophysics}
\vspace{-1.5em}
\author{%
    \begin{minipage}{\textwidth} 
    \centering
    \textbf{Sujit Roy}\textsuperscript{1,2,$\dagger,\ddagger$}, 
    \textbf{Johannes Schmude}\textsuperscript{5,$\dagger,\ddagger$}, 
    \textbf{Rohit Lal}\textsuperscript{1}, 
    \textbf{Vishal Gaur}\textsuperscript{1}, 
    \textbf{Marcus Freitag}\textsuperscript{5}, 
    \textbf{Julian Kuehnert}\textsuperscript{5}, 
    \textbf{Theodore van Kessel}\textsuperscript{5}, 
    \textbf{Dinesha V. Hegde}\textsuperscript{3,4}, 
    \textbf{Andrés Muñoz-Jaramillo}\textsuperscript{7}, 
    \textbf{Johannes Jakubik}\textsuperscript{5}, 
    \textbf{Etienne Vos}\textsuperscript{5}, 
    \textbf{Kshitiz Mandal}\textsuperscript{1}, 
    \textbf{Ata Akbari Asanjan}\textsuperscript{14}, 
    \textbf{Joao Lucas de Sousa Almeida}\textsuperscript{5}, 
    \textbf{Amy Lin}\textsuperscript{1}, 
    \textbf{Talwinder Singh}\textsuperscript{6}, 
    \textbf{Kang Yang}\textsuperscript{6}, 
    \textbf{Chetraj Pandey}\textsuperscript{6}, 
    \textbf{Jinsu Hong}\textsuperscript{6}, 
    \textbf{Berkay Aydin}\textsuperscript{6}, 
    \textbf{Thorsten Kurth}\textsuperscript{15},
    \textbf{Ryan McGranaghan}\textsuperscript{8}, 
    \textbf{Spiridon Kasapis}\textsuperscript{9}, 
    \textbf{Vishal Upendran}\textsuperscript{10}, 
    \textbf{Shah Bahauddin}\textsuperscript{11}, 
    \textbf{Daniel da Silva}\textsuperscript{12}, 
    \textbf{Nikolai V. Pogorelov}\textsuperscript{3,4},
    \textbf{Anne Spalding}\textsuperscript{13}, 
    \textbf{Campbell Watson}\textsuperscript{5}, 
    \textbf{Manil Maskey}\textsuperscript{2},
    \textbf{Madhulika Guhathakurta}\textsuperscript{16},
    \textbf{Juan Bernabe-Moreno}\textsuperscript{5}, 
    \textbf{Rahul Ramachandran}\textsuperscript{2} \\
    \vspace{0.5cm} 
    \textsuperscript{$\dagger$}Equal Contribution, \\
    \textsuperscript{$\ddagger$}\texttt{sujit.roy@nasa.gov}, \texttt{Johannes.Schmude@ibm.com} 
    \thanks{%
    \textsuperscript{1}Earth System Science Center, University of Alabama in Huntsville, AL, USA; 
    \textsuperscript{2}NASA Marshall Space Flight Center, Huntsville, AL, USA; 
    \textsuperscript{3}Department of Space Science, University of Alabama in Huntsville, AL, USA; 
    \textsuperscript{4}Center for Space Plasma and Aeronomic Research (CSPAR), University of Alabama in Huntsville, AL, USA; 
    \textsuperscript{5}IBM Research; 
    \textsuperscript{6}Georgia State University; 
    \textsuperscript{7}Southwest Research Institute; 
    \textsuperscript{8}NASA Jet Propulsion Laboratory; 
    \textsuperscript{9}Princeton University; 
    \textsuperscript{10}SETI Institute; 
    \textsuperscript{11}Laboratory for Atmospheric and Space Physics, University of Colorado Boulder; 
    \textsuperscript{12}NASA Goddard Space Flight Center;
    \textsuperscript{13}Trillium Tech Inc.;
    \textsuperscript{14}Research Institute for Advanced Computer Science, Universities Space Research Association, USA
    \textsuperscript{15}NVIDIA Corp., Santa Clara, USA Caltech, Pasadena, USA;
    \textsuperscript{16}NASA Science Mission Directorate
    }
    \end{minipage}
}

\date{}

\begin{document}

\maketitle

\begin{abstract}
Heliophysics is central to understanding and forecasting space weather events and solar activity. Despite decades of high-resolution observations from the Solar Dynamics Observatory (SDO), most models remain task-specific and constrained by scarce labeled data, limiting their capacity to generalize across solar phenomena. We introduce Surya, a 366M parameters foundation model for heliophysics designed to learn general-purpose solar representations from multi-instrument SDO observations, including eight Atmospheric Imaging Assembly (AIA) channels and five Helioseismic and Magnetic Imager (HMI) products. Surya employs a spatiotemporal transformer architecture with spectral gating and long–short range attention, pretrained on high-resolution solar image forecasting tasks and further optimized through autoregressive rollout tuning. Zero-shot evaluations demonstrate its ability to forecast solar dynamics and flare events, while downstream fine-tuning with parameter-efficient Low-Rank Adaptation (LoRA) adaptation shows strong performance on solar wind forecasting, active region segmentation, solar flare forecasting, and EUV spectra. Surya is the first foundation model in heliophysics that uses time advancement as a pretext task on full-resolution SDO's data.  Its novel architecture and performance suggest that the model is able to learn the underlying physics behind solar evolution.

\end{abstract}

\section{Introduction}

\begin{figure}[htb]
\centering
\includegraphics[width=1\textwidth]{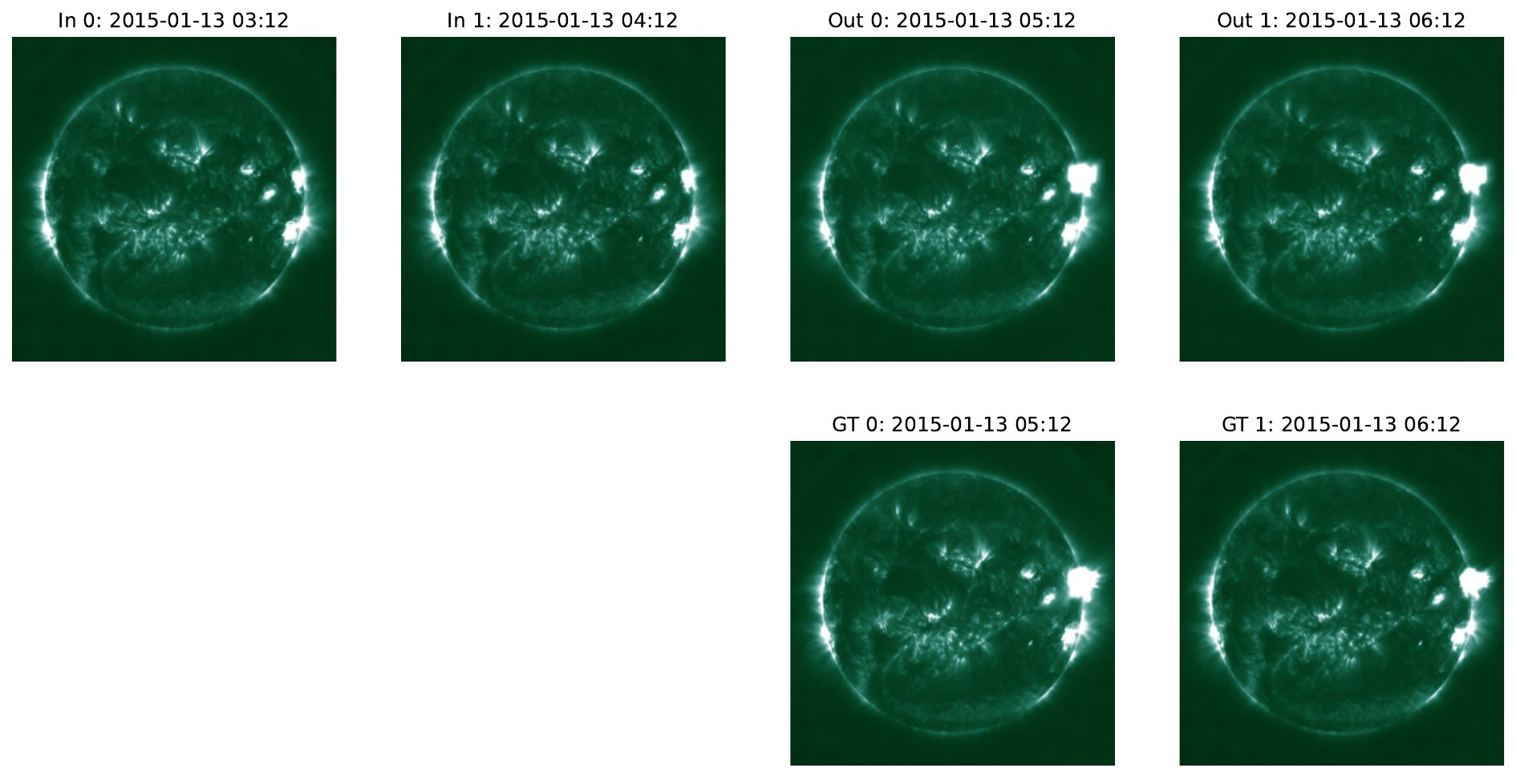} 
\caption{An M-class solar flare that occurred on January 13, 2015, as forecasted by Surya. AIA, 94\AA. Top row, left two columns are model inputs (``In''). Top row, right two columns are model outputs (``Out''). Bottom row shows corresponding ground truth (``GT'').}
\label{fig:solar_flare_visual_20150113}
\end{figure}

% Paragraph 1: general background on how heliophysics impacts our lives (with stakeholders). 
Heliophysics is the study of the Sun and its impact on the solar system. Its relevance goes significantly beyond a purely scientific interest in our star. Indeed, one of the most critical aspects of heliophysics is the study of space weather, which is driven by the Sun's activity, such as solar flares and coronal mass ejections (CMEs). These events can have significant effects on modern technology and consequently society~\citep{Boteler_quebec_2001}. Space weather can disrupt satellite communications and GPS signals, interfere with aviation navigation systems, and degrade the quality of radio transmissions. Strong geomagnetic storms can induce damaging currents in electrical grids, leading to widespread power outages and costly infrastructure repairs~\citep{Oughton_dailyimpactspw_2017}. In space, increased radiation levels pose a severe hazard to astronauts and spacecraft electronics, potentially shortening mission lifespans or endangering human health. Even on Earth, high-latitude flights can be exposed to elevated radiation levels, and critical services that depend on precise timing, such as financial transactions, weather forecasting, and emergency response, can be severely impacted. These impacts create a broad network of stakeholders, including space agencies, satellite operators, aviation authorities, power companies, defense organizations, high-precision farming, and emergency management agencies \cite{Worman2018}. Understanding and predicting solar activity is thus essential for safeguarding critical systems and ensuring resilience to space weather hazards. Beyond its practical importance, heliophysics offers a unique window into fundamental plasma physics in extreme environments and deepens our understanding of how solar activity shapes planetary atmospheres, informing studies of planetary habitability both on Earth and across distant exoplanets \citep{Schrijveretal:2019}.

%current ML models in heliophysics
In recent years, several ML-based methods have been developed to achieve results superior to conventional methods in solving various problems in the field of solar and heliospheric physics. For example, for feature detection on the Sun, be it coronal holes (CHs), active regions(ARs), prominences, or magnetic flux elements \citep{Illarionov18, Mackovjak21, Jiang20, Jarolim21, Baek21, Armstrong19, Inceogluetal:2022}, forecasting solar flares \citep{Ahmed13, Bobra15, Nishizuka17, Nishizuka18, Abed21, Yi21}, solar wind prediction \citep{Upendran20, Brown22, Rajuanddas:2021, Bernoux22}, 3D reconstruction of the coronal magnetic fields \citep{Chifu21, Rahman23}, image to image translation \citep{Szenicer19, DosSantos:2021, Kim19, Jeongetal:2020, Jeongetal:2022}, super-resolution tasks \citep{Diaz18, Dou22}, solar cycle predictions \citep{Okoh18, Benson20, Li21, Prasad22, Bizzarri22}.

% Paragraph 2: 1-2 sentences on foundation models and how they can be used for scientific tasks. 
Foundation models (FMs) are large, pretrained machine learning models that learn general-purpose representations from vast datasets, and they have revolutionized practical applications, most notably in natural language processing and computer vision by capturing rich, transferable features that enable rapid adaptation to diverse downstream tasks with minimal task-specific training. Their potential is now increasingly recognized in scientific disciplines that generate complex, multimodal data, and heliophysics is not an exception \citep{roy2024ai}. Despite decades of extensive solar observations from a fleet of ground-based telescopes \citep{Harvey-etal1996, rimmele2020daniel} and space-based satellites \citep{Pesnell12,domingo1995soho}, current ML applications in heliophysics research often depends on task-specific data \citep{bobra2014helioseismic, kosovich2024time} and models trained from scratch \citep{pandey2021deep, kasapis2024forecasting}, which can be inefficient, prone to overfitting, and limited by the scarcity of labeled data, especially for rare events \cite{Ahmadzadeh2019}, which are often the most interesting ones (for a detailed review \cite{Asensioetal:2023}).

% Paragraph 3: why we need a foundation model in heliophysics. 
A heliophysics FM can address these limitations by learning generalized, physics-aware representations from the wealth of high-resolution, multi-instrument solar data, enabling robust performance across a broad spectrum of predictive, diagnostic, and analytical tasks \cite{Nita2020, Camporeale2020, McGranaghan2024}. By leveraging pre-training on a large database of solar observations, FMs can mitigate the supervision bottleneck, reducing the need for labeled data and improving real-world performance in forecasting rare or extreme events \citep{roy2024ai,walsh2024foundation,majid2024solaris}. Their versatility and adaptability allow fine-tuning for diverse downstream applications, including solar feature detection (e.g., coronal holes \cite{Delouille2018, jarolim_coronalhole}, sunspots~\citep{seth_featureidentification}, active regions \citep{Ji2023,kasapis2023predicting}), transient event forecasting~\citep[e.g. space weather]{Sinha_flares, Hu_dst_2022,upendran_dagger, Georgoulis2024, Ahmadzadeh2021, Nishizuka2018}, and heliospheric modeling \cite{Nguyen2019, Singh2023, SierraPorta2024}, with minimal additional supervision. FMs also offer improved generalization, effectively handling data distribution shifts that hinder traditional models, and their scalability ensures compatibility with increasingly large and high-resolution datasets for multi-scale modeling of solar dynamics. Furthermore, their capacity for multi-modal integration can unlock more accurate and comprehensive predictions of solar activity and heliospheric conditions. Collectively, these capabilities position foundation models as a transformative step toward next-generation data-driven heliophysics.

% Paragraph 4: what type of problems it can solve (transfer learning, class imbalance, ). 
% I think we can use the above para for this

% Paragraph 5: contributions of the paper (introduces a foundation model, shows its effectiveness on various downstream applications)
In this paper, we introduce Surya (Sanskrit for Sun), our heliophysics foundation model trained on multi-channel native-resolution observations from Solar Dynamics Observatory \citep[SDO]{Pesnell12}.
%observations, spanning eight Atmospheric Imaging Assembly (AIA) channels and five Helioseismic and Magnetic Imager (HMI) raster products.
We detail the pretext task design, pretraining protocols, and engineering approaches used to efficiently process and learn from large-scale, multimodal solar data. Our model architecture and training pipeline integrate modern self-supervised learning paradigms to capture high-fidelity, general-purpose solar representations. We demonstrate the model’s versatility across a suite of downstream applications, including active region emergence prediction, active region segmentation, solar flare forecasting, and solar wind forecasting. The evaluations confirm that the model produces consistent, transferable representations that maintain competitive or improved performance across diverse tasks, underscoring its potential as a scalable backbone for both scientific discovery and practical forecasting applications.

\section{Surya FM}

\subsection{SDO Data} \label{sec:dataset}

SDO launched on February 11, 2010, as the first mission of NASA’s Living With a Star Program (LWS). It is dedicated to advancing our understanding of solar variability and its impacts on Earth and near-Earth space. By observing the solar atmosphere at high spatial and temporal resolution across multiple wavelengths, SDO investigates how the Sun’s magnetic field is generated, structured, and released, driving the solar wind, energetic particles, and irradiance variations. These observations aim to enable a predictive capability for solar variations that affect life and technological systems on Earth, collectively known as space weather. 

\begin{figure}[b!]
\centering
\includegraphics[width=1\textwidth]{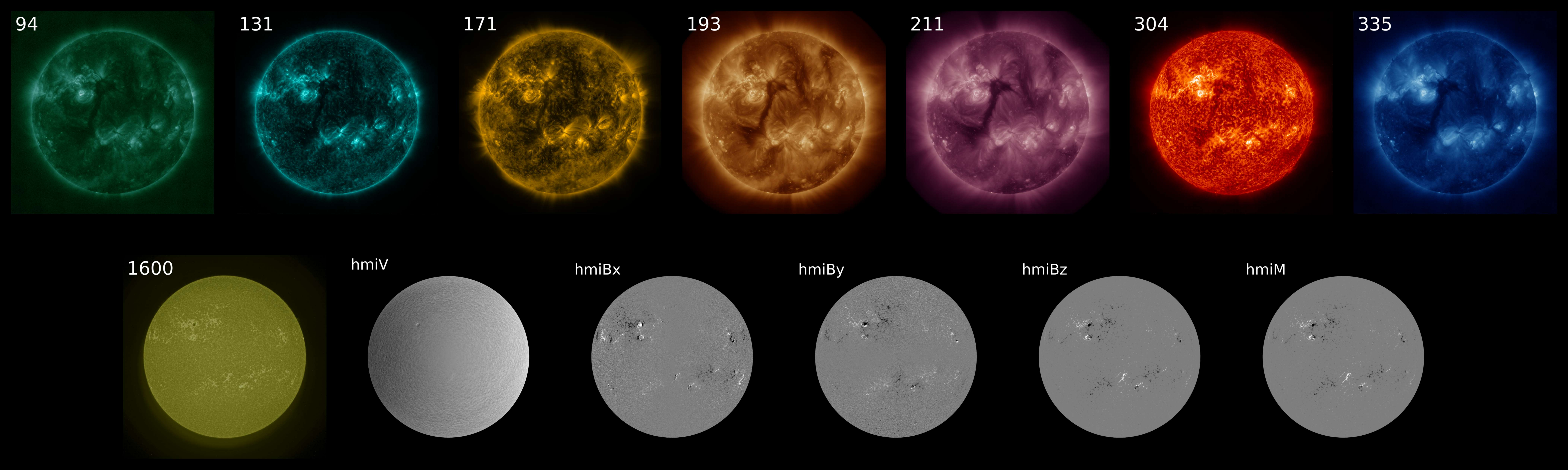} 
\caption{Illustrations of SDO solar imagery used for training Surya: Solar coronal EUV/UV images from AIA (8 channels: 94, 131, 171, 193, 211, 304, 335, 1600~\AA) and solar surface velocity and magnetic field maps from HMI (5 channels: hmiV, hmiBx, hmiBy, hmiBz, hmiM) }
\label{fig:sdo_data}
\end{figure}

For this study, we utilize observations from two primary imaging instruments onboard SDO: the \textit{Atmospheric Imaging Assembly} (AIA)\citep{Lemen12}, which records extreme ultraviolet (EUV) and ultraviolet (UV) photometric intensities, and the \textit{Helioseismic and Magnetic Imager} (HMI)\citep{Schou12}, which provides spectropolarimetric measurements for deriving vector magnetic fields and line-of-sight velocities. Both instruments utilize \(4096 \times 4096\) pixel CCD detectors, capturing full-disk images with sub-arcsecond spatial resolution.

We selected SDO data from AIA and HMI because they uniquely provide 
the highest spatial resolution, full-disk solar observations with high cadence. SDO offers a nearly continuous and uniform dataset spanning 15 years, covering more than a full solar cycle. 
%The simultaneous observations of the Sun by HMI and the multi-wavelength AIA make SDO a natural choice for building a foundation model that leverages spatially resolved and temporally rich solar data to address multiple problems in heliophysics and space weather \citep{roy2024ai}. 
The consistent and simultaneous observations of the Sun by HMI and AIA make SDO a natural choice for building an FM, as its uniquely high spatial and temporal resolution, sustained coverage, and diverse data products provide a rich basis for addressing multiple data-driven learning problems in heliophysics and space weather, while enabling a wide spectrum of science use cases. Furthermore, SDO is an ongoing mission that runs concurrently with the GOES-SUVI operational satellites \citep[which also makes EUV imaging observations][]{Darnel-etal2022D} and ground-based magnetogram and Doppler measurements by the GONG network \citep{Harvey-etal1996}.  This ensures that the Surya foundation model will remain applicable to future unseen data, thereby extending its utility beyond historical records.

\subsubsection{Dataset Preparation}

We construct our dataset by homogenizing
%integrating
multi-channel AIA EUV/UV imagery from the \texttt{aia.lev1\_euv\_12s} and \texttt{aia.lev1\_uv\_24s} series with HMI products from the \texttt{hmi.M\_720s} (line-of-sight), \texttt{hmi.B\_720s} (vector), and \texttt{hmi.V\_720s} (line-of-sight) series. All data are downloaded from the Joint Science Operations Center (JSOC\footnote{\url{http://jsoc.stanford.edu}}). The temporal cadence is standardized to 12 minutes, matching the HMI vector magnetic field data series, with AIA observations temporally co-aligned within \(\pm 2\) minutes where possible and subject to quality flag checks. Table~\ref{tab:sdo_properties} summarizes the properties of the individual channels used in this study, including their instrument-specific temporal and spatial resolutions as well as the dynamic range of the measurements.

The Level~1.0 AIA data are preprocessed by updating spacecraft pointing, 
aligning the y-axis of the image with solar north, rescaling to a uniform 
$0.6\,\dprime/\mathrm{pixel}$ grid, bringing solar disk center to the image center, and normalizing for exposure-time variability. Time-dependent instrument degradations are corrected using calibration factors, with values clamped to the instrument’s dynamic range to avoid saturation artifacts. The Level~1.5 HMI data, with a pixel scale of $0.5\,\dprime/\mathrm{pixel}$, are preprocessed to align with the preprocessed  AIA images at a $0.6\,\dprime/\mathrm{pixel}$ image scale by reprojecting using bilinear interpolation, which also corrects for the HMI roll angle. Finally, the solar disk radius in both datasets is fixed to a constant value to remove the nonphysical variations in the apparent solar disk radius due to the elliptical orbit of the Earth.

The resulting database consists of spatially and temporally aligned, preprocessed, multi-wavelength solar images that are well-suited for machine learning applications. This harmonized processing pipeline ensures that variations within the dataset are predominantly physical in origin rather than instrumental or geometric, thereby enabling robust downstream applications in heliophysics and space weather. Figure~\ref{fig:sdo_data} presents an example set of preprocessed images from the eight AIA and five HMI channels used in this study.

\begin{table}[ht]
  \centering
  \caption{Key properties of the SDO/AIA and SDO/HMI instruments. Cadence values refer to both instrument-native and standardized dataset cadence.} \vspace{0.2cm}
  \resizebox{\linewidth}{!}{%
  \begin{tabular}{lcccc}
    \toprule
    \textbf{Instrument} & \textbf{Resolution} & \textbf{Cadence (instr./dataset)} & \textbf{Dynamic range} & \textbf{Channels / Measurements} \\
    \midrule
    AIA & $1.2^{\prime\prime}$ ($\approx$725 km) & 12 s, 24 s / 12 min & 0--16{,}383 DN & 94, 131, 171, 193, 211, 304, 335, 1600~\AA \\
    HMI & $1.0^{\prime\prime}$ ($\approx$870 km) & 45 s, 12 min / 12 min & $\pm 4{,}500$ G (B), $\pm 10^{4}$ m/s (V) & $B_x$, $B_y$, $B_z$, $B_{\mathrm{los}}$, $V_{\mathrm{los}}$ \\
    \bottomrule
  \end{tabular}%
  }
  \label{tab:sdo_properties}
\end{table}

\subsubsection{Dataset Statistics}
\label{sec:Stats}

Our database contains ML-ready SDO data captured from May 13, 2010, to December 31, 2024. During this interval, there are about 2.9\% data unavailable (18,261 out of 623,280 total timestamps) for \textit{hmi.M\_720s} series. The processed level 1.5 AIA and HMI data are stored in netCDF files (float32 format), with one file per cadence. Each file contains data from five 12-minute timesteps within that hour, with the data shape of [13, 4096, 4096]. Each netCDF file is using lz4 compression for better training performance ($\approx$30\% improvement) and is about 630 MB. The total size of the data for dataset is around 257 TB.

\paragraph{Data details}
We define a solar observation as a multi-channel, co-registered raster representing simultaneous measurements from AIA and HMI onboard SDO. Each observation is encoded as a three-dimensional tensor:
$X \in \mathcal{R}^{13 \times 4096 \times 4096} $\\
Where:
\begin{itemize}
    \item The first dimension indexes the %spectral or 
    physical measurement channel,
    \item The second and third dimensions represent the spatial domain, corresponding to the solar disk with a native resolution of $0.6\,\dprime/\mathrm{pixel}$.
    %approximately 1 arcsecond per pixel.
\end{itemize}

Channel Composition Information:
The 13 channels are composed as follows:
\begin{itemize}
    \item 8 AIA Channels: 7 EUV (94 Å, 131 Å, 171 Å, 193 Å, 211 Å, 304 Å, 335 Å) and 1 UV (1600 Å).
    %: 94 Å, 131 Å, 171 Å, 193 Å, 211 Å, 304 Å, 335 Å, 1700 Å
    \item 5 HMI Channels: 1 LOS magnetic field map B\_{los}, 3 vector magnetic field component maps (B\_x, B\_y, B\_z), and 1 Doppler velocity map V\_{los}.
\end{itemize}
Each AIA channel provides EUV or UV intensity measurements of the solar atmosphere, capturing coronal and chromospheric emission at distinct temperature regimes. HMI-derived channels capture photospheric vector magnetic field observations and line-of-sight plasma motion.

\paragraph{Train/test split}
\label{sec:train_test_split}

For the train–test partition, observations from 2011 to 2019 were segmented by day-of-year. Days 1–14 and 32–45 of each year were excluded as temporal buffers to mitigate potential information leakage due to short-term temporal autocorrelation in solar activity. The interval spanning days 15–31 of each year was reserved exclusively for the test set, while all remaining days from day 46 onward were assigned to the training set. While data for Solar Cycle 25 is available, we excluded it to enable further validation for both extended pretext tasks and downstream applications where there is potential for implicit data leakage.

\paragraph{Normalization}

\begin{figure}[H]
\centering
\includegraphics[width=0.8\textwidth]{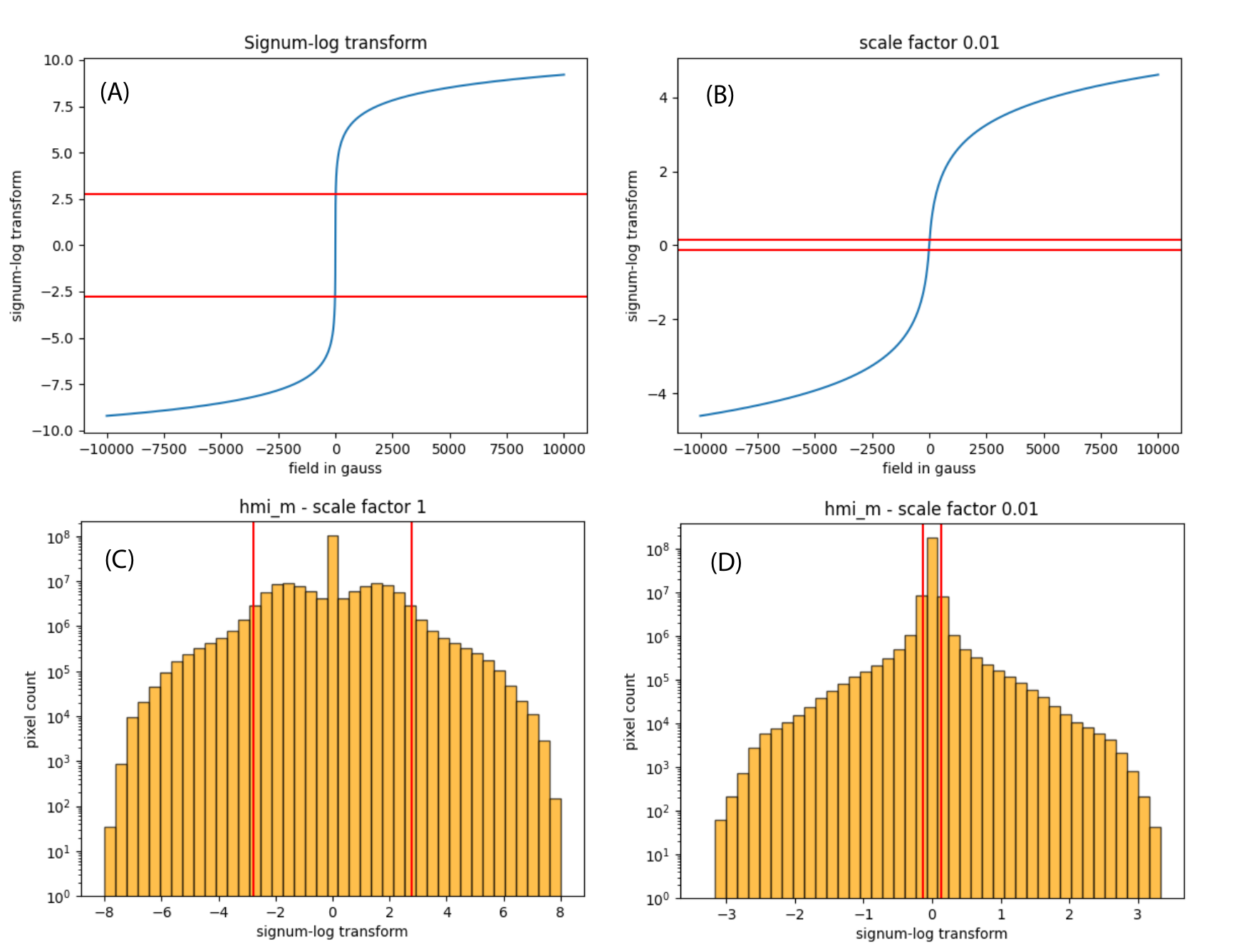}
\caption{Scaled signum-log transform reduces the range over which noise dominates. (A) Standard signum-log transform. (B) Scaled signum-log transform with scale factor $10^{-2}$. (C) Histogram of HMI\_m channel values after signum-log transform. (D) Histogram of HMI\_m channel values after scaled signum-log transform with scale factor $10^{-2}$. Red lines indicate the limits of the range where the magnetic field magnitude is smaller than the noise level of 15 Gauss.}
\label{fig:signum_log}
\end{figure}

SDO data is highly skewed due to the predominance of pixels depicting the quiet sun and the presence of a small number of pixels depicting extreme events such as solar flares. 
%or coronal mass ejections. 
Modeling such data often benefits from log-transformation before applying the standard scaling. Specifically, we consider the signum-log transform $\sign(x) \times \log1p(\vert \mathbf{X} \vert)$, which can be applied equally to the strictly positive AIA channels and the HMI channels that contain both positive and negative values (where $log1p(x)$ is the natural logarithm of $1+x$). However, as figure \ref{fig:signum_log} demonstrates, the standard signum-log transform comes with the downside that it stretches the inherently noisy low-magnetic field range of HMI channels at the expense of data points with good signal-to-noise ratio. Given the inherent noise level of the HMI\_m channel of 15 Gauss, about 1/3 of the transformed scale would be occupied by data that is dominated by the inherent detector noise, while the dynamic range of extreme events may be compressed too much. As a compromise between applying the full signum-log or no transformation at all, we propose applying a scale factor of $10^{-2}$ to the raw data before applying the signum-log. Following this transformation, we standard-scale the data per channel. Thus, the complete transformation per channel is
\begin{equation}\label{eq:signum_log_scaling}
    \frac{\sign(\mathbf{X}) \log1p(\vert 10^{-2} x \vert) - \mu}{\sigma + \epsilon}.
\end{equation}
Here, $\mu$ and $\sigma$ are the per channel global means and standard deviations following the signum-log transformation; $\epsilon$ is a small constant, we use $10^{-8}$. A histogram of the resulting data distribution in \ref{fig:signum_log} (D) shows that the inherently noisy values are now occupying a much narrower range and the model can focus on learning the important intermediate and extreme activity ranges of the sun.
\subsection{Pretext task and baselines}

\subsubsection{Predicting future SDO imagery as a pretext task}
\label{sec:forecasting_as_pretext}

Prior to considering AI architectures, one needs to decide on a pretext task. Here one can take inspiration from prior work in computer vision, but also other scientific domains such as earth observation or atmospheric physics. Although there is a myriad of pretext tasks (i.e.~self-supervised training methodologies) the most dominant ones are arguably masked reconstruction \citep{he2022masked}, contrastive objectives \citep{chen2020simple}, and finally autoencoders. In addition, recent work in atmospheric physics moreover shows that AI models can learn complex temporal dynamics purely from data by regressing onto a future time step \citep{pathak2022fourcastnet, lam2023learning}. On the other hand, masked reconstruction has been successfully used in earth observation \citep{jakubik2023foundation, szwarcman2024prithvi} while \cite{schmude2024prithvi} used a mixed objective combining reconstruction with forecasting. Yet in the end we note that our downstream tasks are frequently in nature and choose forecasting as a pretrainig objective. In heliophysics, \cite{majid2024solaris} trained with a 12-hour ahead forecasting objective while \cite{walsh2024foundation} explored both MAEs and autoencoders. Both papers did so at 512 by 512 pixel resolution.

For future work, let us point out that a band-to-band translation pretext task, as was used with great success by \cite{jakubik2025terramind} in earth observation, might be a strong alternative if combined with a temporal objective. Especially given the inherent multi-modality of SDO data.

In either case, we use two timestamps, 60 minutes apart, as input to the model and train the model by regressing on SDO data 60 minutes in the future. As is now standard in models for atmospheric physics, this is followed by a second phase of pretraining where the outputs of Surya are used as inputs to predict 120, 180, \dots minutes into the future. This is generally referred to as ``autoregressive rollout tuning''. During rollout tuning, we average the loss across all steps. See section \ref{sec:pretraining_protocol} for details on our pretraining protocol.

To formalize this, we denote observed frames as $\mathbf{X}_t$ and the model as $f_\theta$. Then we train Surya with a mean square error (MSE) objective as follows:
% \begin{equation}
%     \left\lbrack \mathbf{X}_{t+1} - f_\theta\left( \mathbf{X}_t, \mathbf{X}_{t-1} \right) \right\rbrack^2.
% \end{equation}
\begin{equation}
    \mathcal{L}_{\text{MSE}} \;=\; \frac{1}{N} \left\lVert \mathbf{X}_{t+1} - f_\theta\!\left( \mathbf{X}_t, \mathbf{X}_{t-1} \right) \right\rVert^2,%_2^2,
\end{equation}

If we further denote model output as $\hat{\mathbf{X}}_{t+1} = f_\theta ( \mathbf{X}_t, \mathbf{X}_{t-1} )$, autoregressive prediction takes the form
\begin{equation}
    \hat{\mathbf{X}}_{t+2} = f_\theta\left( \hat{\mathbf{X}}_{t+1}, \mathbf{X}_t \right) = f_\theta \left( f_\theta \left( \mathbf{X}_t, \mathbf{X}_{t-1} \right), \mathbf{X}_t \right).
\end{equation}
The rollout loss is then
% \begin{equation}
%     \frac{\left\lbrack \mathbf{X}_{t+1} - f_\theta\left( \mathbf{X}_t, \mathbf{X}_{t-1} \right) \right\rbrack^2
%     + \left\lbrack \mathbf{X}_{t+2} - f_\theta \left( f_\theta \left( \mathbf{X}_t, \mathbf{X}_{t-1} \right), \mathbf{X}_t \right) \right\rbrack^2}{2}
% \end{equation}

\begin{equation}
    \mathcal{L}_{\text{rollout}} \;=\; 
    \frac{1}{2N} \Bigg(
        \left\lVert \mathbf{X}_{t+1} - f_\theta\!\left( \mathbf{X}_t, \mathbf{X}_{t-1} \right) \right\rVert^2%_2^2
        + \left\lVert \mathbf{X}_{t+2} - f_\theta \!\left( f_\theta \!\left( \mathbf{X}_t, \mathbf{X}_{t-1} \right), \mathbf{X}_t \right) \right\rVert^2%_2^2
    \Bigg),
\end{equation}
and equivalent for multi-step rollouts. At inference time, this scheme enables longer forecasts from two initial observations an hour apart.

\subsubsection{Baseline scores}

With the pretraining task decided, it is important to identify a series of baseline scores to compare against. The purpose is primarily to put model losses into context during training and development. The first of these is simply persistence. Using the notation from section \ref{sec:forecasting_as_pretext}, the persistence forecast is simply $\hat{\mathbf{X}}_{t+1} = \mathbf{X}_t$. Technically speaking, note that one can obtain an improved persistence forecast by averaging over multiple timestamps along the lines of $\hat{\mathbf{X}}_{t+1} = \left( \mathbf{X}_t + \mathbf{X}_{t-1} \right) / 2$. The reason for this is that the averaging procedure smooths out sharp features. Especially due to the sun's rotation, sharp features lead to a double penalty on MSE loss of persistence. However, our persistence scores serve simply as a baseline, so we use $\hat{\mathbf{X}}_{t+1} = \mathbf{X}_t$.

\begin{table}
    \centering
    \caption{Baseline scores for 1 hour ahead forecasting. In model units.}
    \label{table:Baseline scores}
    \begin{tabular}{lrr}
        \toprule
        \textbf{Baseline} & \textbf{Parameters} & \textbf{Loss (MSE)} \\
        \midrule
        Persistence & N/A & $0.594044030$ \\
        Learned flow & $642$ & $0.337624282$ \\
        \bottomrule
    \end{tabular}
\end{table}

The other relevant baseline is given by solar rotation. Rather than hard-coding rotation via a known equation, we decide to learn the effect from data. To do so, we assign coordinates $\lbrack -1, 1 \rbrack \times \lbrack -1, 1 \rbrack$ to the 4096 x 4096 pixels. That is, the bottom left pixel is $(-1, -1)$, the top right pixel $(1, 1)$ etc. Then we train a very small Multi-Layer Perceptron (MLP) that takes these coordinates of each pixel as input and yields a vector as output. The result is a learned vector field along which we interpolate the data. As the input is data independent, this trains a constant flow-field along which the data is moved to effectively learn the rotation. If we denote the coordinates of each pixel as $\mathbf{x}$, the MLP as $M_\theta$, and the operation to interpolate\footnote{Implementation wise, we use the \texttt{F.grid\_sample} method from PyTorch for interpolation. I.e.~\texttt{F.grid\_sample(x, flow\_field, mode="bilinear")}, where \texttt{x} is the data $\mathbf{X}$ and \texttt{flow\_field} the output of the MLP $M_\theta$.} along a vector as $\nearrow$, we train this baseline as follows:
\begin{equation}
    \left\lbrack \mathbf{X}_{t+1} - M_\theta\left(\mathbf{x}\right) \nearrow \mathbf{X}_t \right\rbrack^2.
\end{equation}
Our MLP consists of two linear layers and an internal dimension of 128 with GELU activation. Thus, it has a total of 642 parameters. Table \ref{table:Baseline scores} shows the baseline scores in the model units of equation \eqref{eq:signum_log_scaling}.

\subsection{Architecture}

% \subsubsection{Spectformer with LS Attention}

\begin{figure}[!ht]
\centering
\includegraphics[width=1\textwidth]{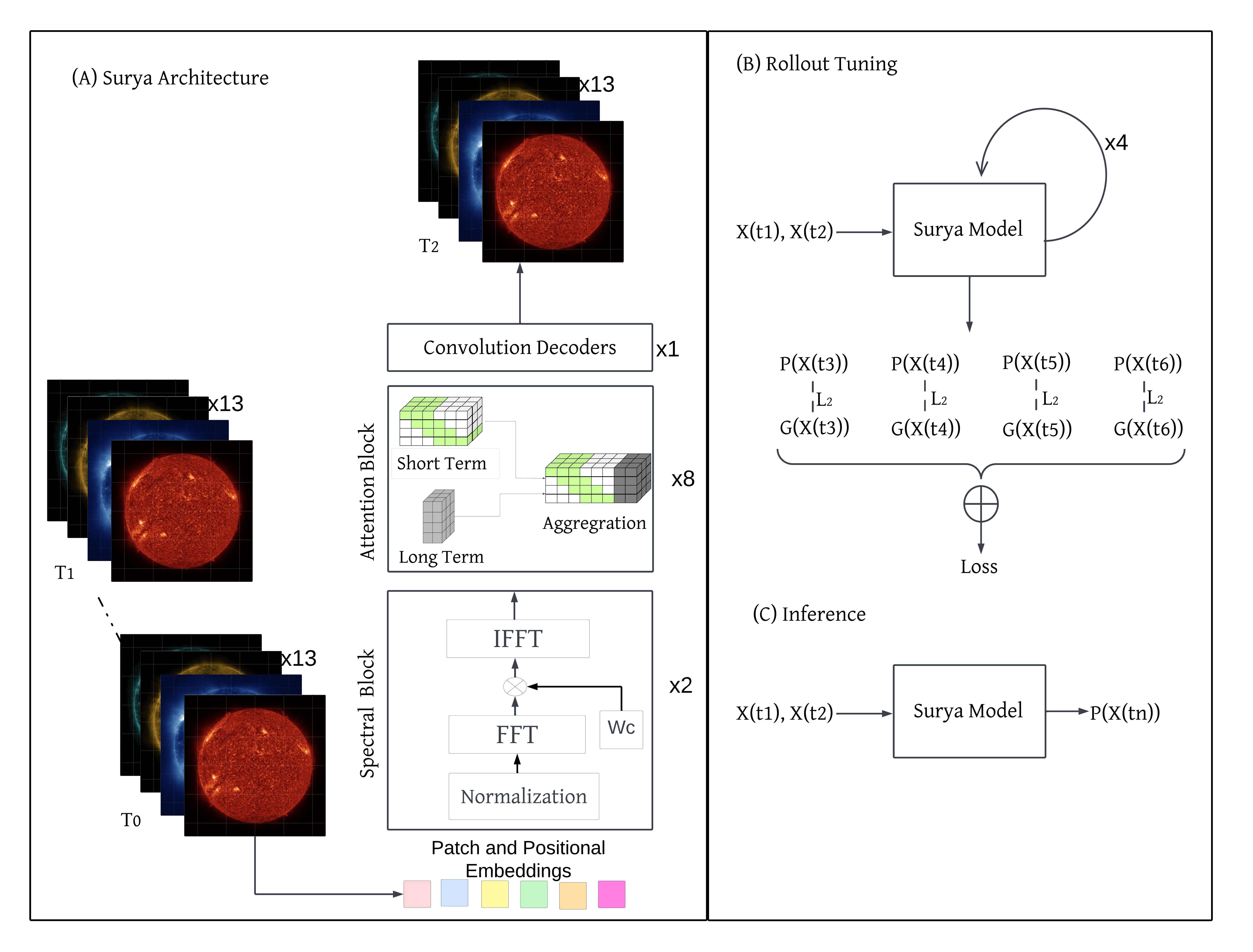} 
\caption{(A). The architecture of the Surya Foundation Model which uses 2 spectral blocks, 8 attention blocks based on long-short attention, and 1 decoder block. We are using a learnable weight parameter ($W_c$) after performing $FFT$ for estimating the weights of frequency components. We then perform an inverse $FFT$ to transform the information back into physical space, as described in \cite{patro2025spectformer}. The attention head is designed on the principle of long-short attention \cite{zhu2021long}, where we calculate short-term attention by sliding window, and long-range attention by dynamic projections. (B). Overview of 4-step rollout finetuning, where the model is given a 2-timestep input and then predicts the next 4 steps autoregressively. C. Inference of the model with 2 time step input and an hour forecast.}
\label{fig:lst_spect}
\end{figure}

% \paragraph{Long-short Attention:}
% To have larger number of sequence tokens, we came up with another strategy of mixing tokens and removing regular attention with the approach proposed by \citep{long_short_term_transformer}. In this approach, dynamic projection-based long-range attention is combined with local windowed short-term attention to effectively capture both global and local features in the input data. Specifically, long-range attention reduces the dimensionality of the Key and Value embeddings by projecting them into a smaller set of tokens using a projection matrix derived from the original Key embeddings. This mechanism allows the model to efficiently manage distant correlations across the sequence. On the other hand, short-term attention divides the input sequences into non-overlapping segments, focusing on capturing fine-grained local correlations. By merging these two mechanisms—long-range dynamic projection for distant dependencies and short-term attention for detailed local interactions—the model successfully addresses different scales of feature representation. By utilizing this approach we were able to train the model with a $16 \times 16$ patch size with a batch size of $4$ on A100 $80$ GB GPUs. Figure \ref{fig:lst_spect} shows the modified architecture of the model. In this version of model, we trained on embedding space of $1024$ along with 12 layers.
\subsubsection{Overview of Surya Model Architecture}

The Surya (figure \ref{fig:lst_spect}) is a 2-D transformer architecture for high-resolution forecasting of SDO imagery and solar dynamics. It integrates frequency-domain filtering with efficient multi-scale attention to capture both fine-scale and global spatio-temporal
%temporal-spatial 
dependencies. The architecture, shown in fig.~\ref{fig:lst_spect}, consists of two spectral gating blocks, eight long-short attention blocks, and a decoder block for reconstruction in the physical domain. See section \ref{sec:ablation_studies} for ablation studies.

\paragraph{Tokenization} The raw input data to Surya is SDO data from 13 different channels scaled according to \eqref{eq:signum_log_scaling}. Using two timestamps, the input data initially has the shape $13 \times 2 \times 4096 \times 4096$. To tokenize the data, we simply flatten the channel and temporal dimensions and use a simple linear layer. The internal dimension is $D=1280$. Given a patch size of $16 \times 16$,
%16 by 16, 
we end up with $N=65,536$ tokens:
\begin{equation}
\begin{aligned}
    C \times T \times H \times W &\mapsto N \times D \\
    \left(13 \times 2\right) \times \left(256 \times 16\right) \times \left(256 \times 16\right) &\to 65,536 \times 1,280.
\end{aligned}
\end{equation}
Surya uses a Fourier position embedding.

\paragraph{Spectral Gating Blocks}

Let $\mathbf{X} \in \mathbb{R}^{B\times N\times D}$ denote the embedded spatiotemporal tokens, where $B$ is the batch size. Each spectral block reshapes $\mathbf{X}$ into $\mathbf{X}_s \in \mathbb{R}^{B\times H_p\times W_p\times D}$, where $H_p = W_p = 256$ are height and width in token space, and applies a 2-D real Fast Fourier Transform (rFFT):
\begin{equation}
\widetilde{\mathbf{X}} = \mathcal{F}(\mathbf{X}_s),
\end{equation}
followed by modulation with a learnable complex-valued weight $W_c \in \mathbb{C}^{H_f\times W_f\times D}$:
\begin{equation}
\widetilde{\mathbf{X}}^\prime = \widetilde{\mathbf{X}} \odot W_c.
\end{equation}
Here, $H_f=256$, $W_f=129$, and $\odot$ denotes element-wise complex multiplication.\footnote{The weight matrices $W_c$ are large at $84,541,440$ real parameters. See section \ref{sec:ablation_studies} for a discussion. Note also that the imbalance between $H_f$ and $W_f$ arises from $\mathbf{X}$ being real and the use of the real FFT here.} This adaptively re-weights frequency components to emphasize informative spectral bands and suppress noise. The result is transformed back to the physical domain via inverse rFFT:
\begin{equation}
\mathbf{X}^\prime_s = \mathcal{F}^{-1}(\widetilde{\mathbf{X}}^\prime),
\end{equation}
and refined through a residual connection and feed-forward network (FFN):
\begin{equation}
\mathbf{X}_{\ell+1} = \mathbf{X}_\ell + \mathrm{FFN}\!\left( \mathrm{LN}\!\left( \mathbf{X}' \right) \right).
\end{equation}
$\mathrm{LN}(\cdot)$ denotes Layer Normalization and $\mathbf{X}^\prime$ is $\mathbf{X}^\prime_s$ reshaped back to sequence order.

\paragraph{Long-Short Attention Blocks}

The attention backbone consists of $L=8$ layers that fuse \textit{local} and \textit{global} attention pathways, following the principle of the Long--Short Transformer~\cite{zhu2021long} adapted for 2-D spatiotemporal tokens.

Local (short-range) attention operates within non-overlapping spatial windows of size $w\times w$, capturing fine-scale dependencies. For each window $\Omega$, queries, keys, and values are restricted to $\Omega$, yielding:
\begin{equation}
\mathrm{Attn}_{\mathrm{short}}(\mathbf{X}) = \mathrm{Softmax}\!\left( \frac{Q_\Omega K_\Omega^\top}{\sqrt{d_k}} + \Delta_{\mathrm{rpe}} \right) V_\Omega,
\end{equation}
where $\Delta_{\mathrm{rpe}}$ is an optional relative positional bias, and $d_k$ is the per-head dimension.

Global (long-range) attention uses dynamic low-rank projection: keys and values are content-adaptively compressed into a rank-$r$ basis,
\begin{equation}
\overline{K} = \alpha K, \quad \overline{V} = \alpha V,
\end{equation}
where $\alpha \in \mathbb{R}^{B\times h\times r\times N}$ is a learned mixing weight obtained from the keys. Queries then attend to $(\overline{K},\overline{V})$ across the entire sequence:
\begin{equation}
\mathrm{Attn}_{\mathrm{long}}(\mathbf{X}) = \mathrm{Softmax}\!\left( \frac{Q \overline{K}^\top}{\sqrt{d_k}} \right) \overline{V}.
\end{equation}

The two branches are normalized to align their scales and then concatenated along the key-value dimension:
\begin{equation}
\mathbf{X}^\prime = \mathrm{Concat}\!\left( \mathrm{Attn}_{\mathrm{long}}, \mathrm{Attn}_{\mathrm{short}} \right),
\end{equation}
followed by a residual MLP:
\begin{equation}
\mathbf{X}_{\ell+1} = \mathbf{X}_\ell + \mathrm{MLP}\!\left( \mathrm{LN}\!\left( \mathbf{X}^\prime \right) \right).
\end{equation}
This design efficiently combines localized modeling with global context aggregation, achieving multi-scale representation learning at reduced complexity as proposed by \cite{zhu2021long}.

\paragraph{Decoder}

The decoder is a lightweight projection that maps the final token representation $\mathbf{X}_L$ back to the spatial domain:
\begin{equation}
\mathbf{Y} \in \mathbb{R}^{B\times C\times H\times W} = \mathrm{Unembed}(\mathbf{X}_L),
\end{equation}
where $C$ is the number of output channels.

\subsection{Pretraining}

\subsubsection{Scaling}

In its final configuration, Surya comprises 366 million parameters. Considering that two timestamps of SDO data comprise
\begin{equation}
     2 \times 13 \times 4096 \times 4096 \times 32 \text{bit} = 1.7 \text{GB}
\end{equation}
of data, GPU memory is a primary concern. At a patch size of $16 \times 16$, we are dealing with $65,536$ tokens. To deal with memory pressure, we use Fully Sharded Data Parallel (FSDP), mixed precision as well as gradient checkpointing. The model's input and output layers operate in \texttt{float32}, yet the transformer layer uses \texttt{bfloat16}. Note that the spectral gating layers explicitly cast to \texttt{float32} for the FFT operations. Table \ref{table:Ablation studies} shows 1 hour ahead performance and memory consumptions of Surya as well as ablations and baselines. Here, the ``No spectral gating'' ablation replaces the two spectral gating blocks with additional long-short attention. As the table shows, the use of the spectral gating layers leads to the same loss at 6\% less GPU memory.

\subsubsection{Pretraining protocol}
\label{sec:pretraining_protocol}

As outlined above, pretraining of Surya followed what has become a standard approach in AI forecasting models in atmospheric physics \citep{pathak2022fourcastnet,lam2023learning}. That is, Surya was trained with a two phase approach consisting of one step ahead forecasting and subsequent rollout tuning.

In phase one, we trained Surya for 160,000 gradient descent steps on 128 NVIDIA A100 GPUs. The model is trained with batch size 1 (per GPU), making an effective batch size of 128. Throughout training, we use cosine annealing to modify the learning rate from $10^{-4}$ to $10^{-5}$. Note that we did not find a need for a dedicated warm-up period to stabilize training. We clipped gradients at $0.1$. We use the \texttt{AdamW} optimizer from PyTorch with default values for its parameters \texttt{betas}, \texttt{eps}, and \texttt{weight\_decay}.

The rollout tuning phase imposes additional demands on GPU memory. We now use gradient checkpointing after all ten spectral and transformer layers. First, we train two steps ahead -- i.e., the initial step plus one autoregressive step -- for 20,000 gradient descent steps at a constant learning rate of $10^{-5}$. Then we subsequently train three, four, and five steps ahead for 4,000 steps each at a learning rate of $10^{-6}$. This unbalanced schedule was partially motivated by the fact that a single gradient descent step takes longer and longer lead times. At the same time, loss curves showed continuous improvement for the 2 step ahead case until step 20,000. Generally, we used 64 GPUs for rollout tuning.

\subsection{Zero-shot Evaluation}

\subsubsection{Predicting future SDO data (forecasting)}

\begin{figure}[htb]
\centering
\includegraphics[width=0.8\textwidth]{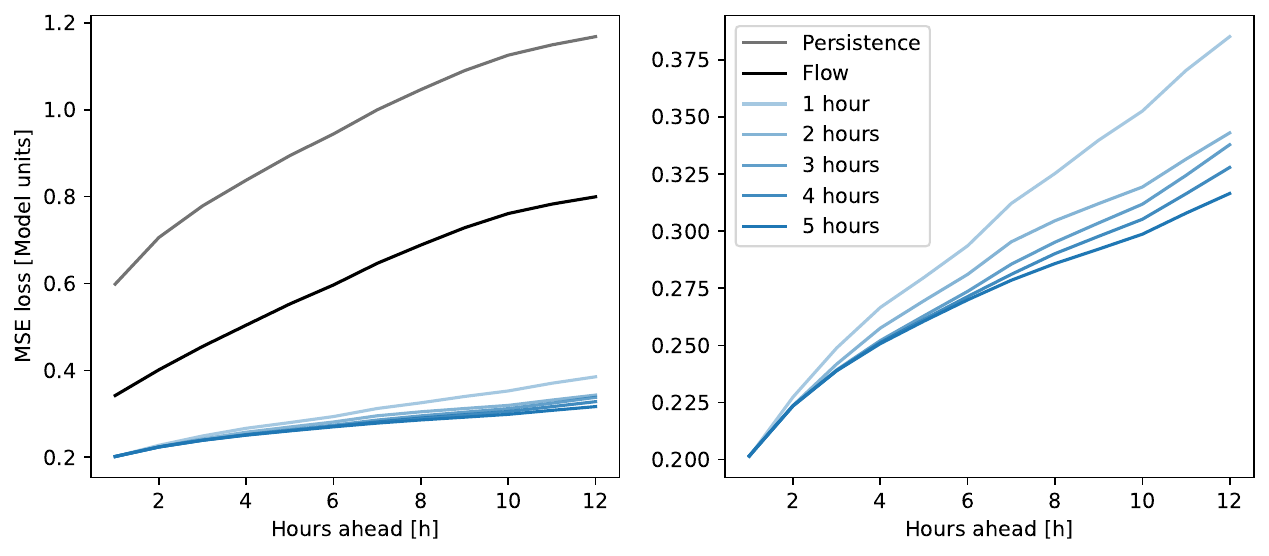} 
\caption{Forecasting performance of Surya. MSE across all channels in model units. Left hand side shows persistence and the flow model used to capture solar rotation. Both sides show the output of phase 1 of pretraining (``1 hour'') as well as various stages of rollout tuning: 2 hour ahead rollout tuning to 5 hour ahead rollout tuning. The latter of these is the last stage of pretraining and thus shows the zero-shot performance of Surya.}
\label{fig:rollout_eval}
\end{figure}

We can get a first understanding of Surya's capabilities, as well as the effectiveness of the pretraining protocol, by predicting future SDO images. The result of this evaluation can be seen in Figure~\ref{fig:rollout_eval}. Rollout tuning improves the performance at 12 hours ahead by $10.9$, $12.3$, $14.9$, and $17.8$\% respectively when compared against the state of the model after phase 1. At the longer lead times, the performance improvements of Surya due to rollout tuning have not yet reached saturation. From a technical perspective, Surya could be tuned up to 24 hours ahead with no modifications to the code when using an 80 GB A100 GPU. The limiting factor here was actually data loading speed with GPUs waiting for data. It might be worthwhile to do additional rollout tuning with longer lead times in the future.

\subsubsection{Visual prediction of solar flares}

\begin{figure}[htb]
\centering
\includegraphics[width=1\textwidth]{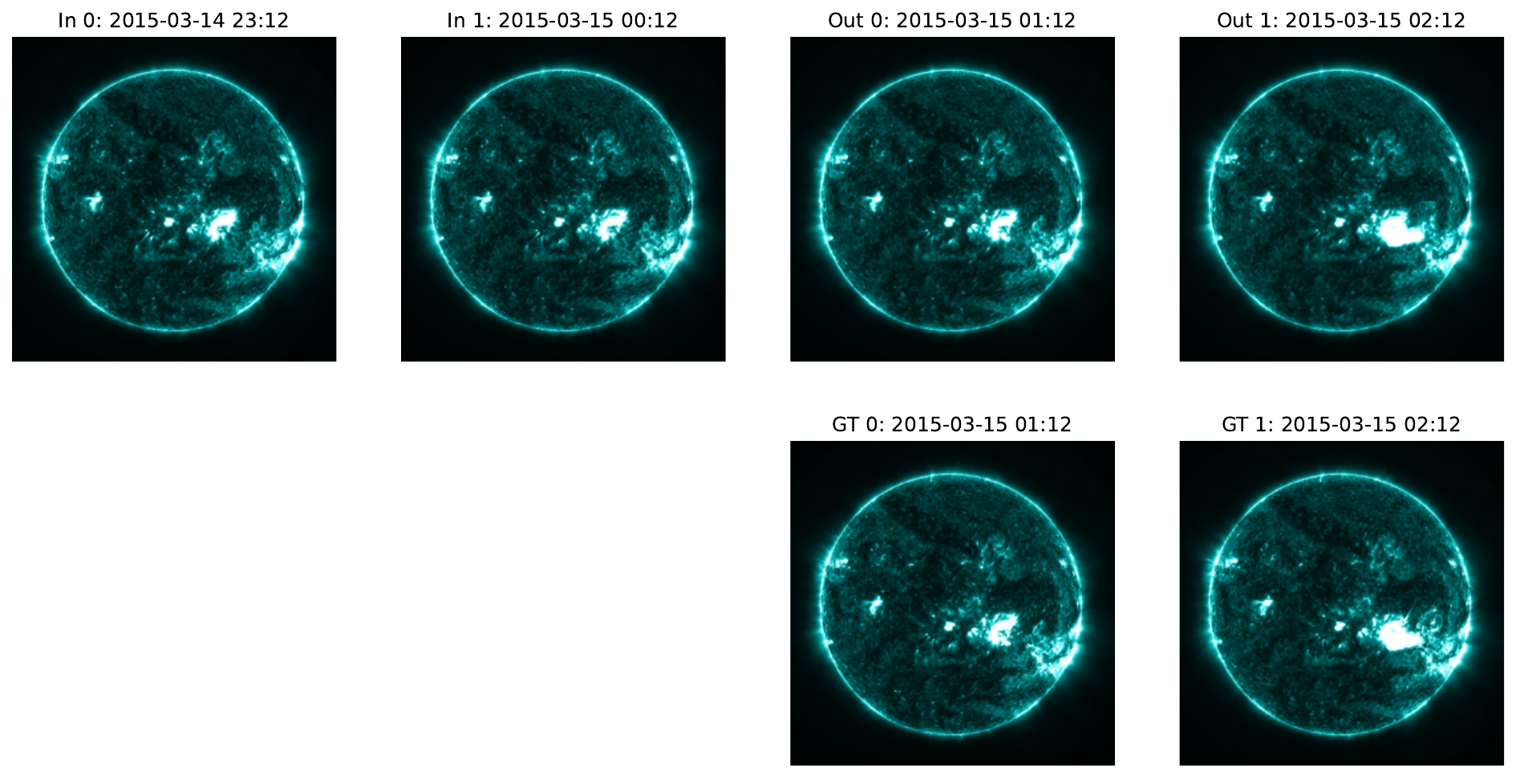} 
\caption{St.~Patrick's Day Event as modeled by Surya. AIA, 131\AA. Top row, left two columns are model inputs (``In''). Top row, right two columns are model outputs (``Out''). Bottom row shows corresponding ground truth (``GT'').}
\label{fig:solar_flare_visual_20150315}
\end{figure}

To further test and evaluate Surya's capabilities to predict future SDO images -- and thus the future state of the sun -- we initialize Surya just prior to a solar flare event. To start, we do this on training data, i.e., imagery that Surya saw during its pretraining period. Figure \ref{fig:solar_flare_visual_20150315} shows this for the St.~Patrick's Day Event \cite{wu2016first}. 
%As reported in \cite{wu2016first}, ``SOHO/LASCO C2 [\dots] recorded a CME [\dots] that erupted from the southwest at 01:48 UT.'' 
Compare the model output of Surya shown in figure \ref{fig:solar_flare_visual_20150315}. It exhibits a clear evolution of model output between 01:12 and 02:12 UTC, fitting the actual observation.

%Naturally, this is training data. As a matter of fact, it stands to reason that a model pretrained with an MSE objective will at very least \emph{remember} an event such as that of figure \ref{fig:solar_flare_visual_20150315}.
Naturally, this is part of the training data, and one would expect a model pretrained with an MSE objective to at least \emph{remember} an event such as that illustrated in Figure~\ref{fig:solar_flare_visual_20150315}. After all, the flare is a clearly visible feature with much brighter pixels than the rest of the sun, having a strong impact on the MSE score. Thus, the critical question is whether we can make similar observations on testing data.

\begin{figure}[hp]
\centering
\includegraphics[height=0.9\textheight]{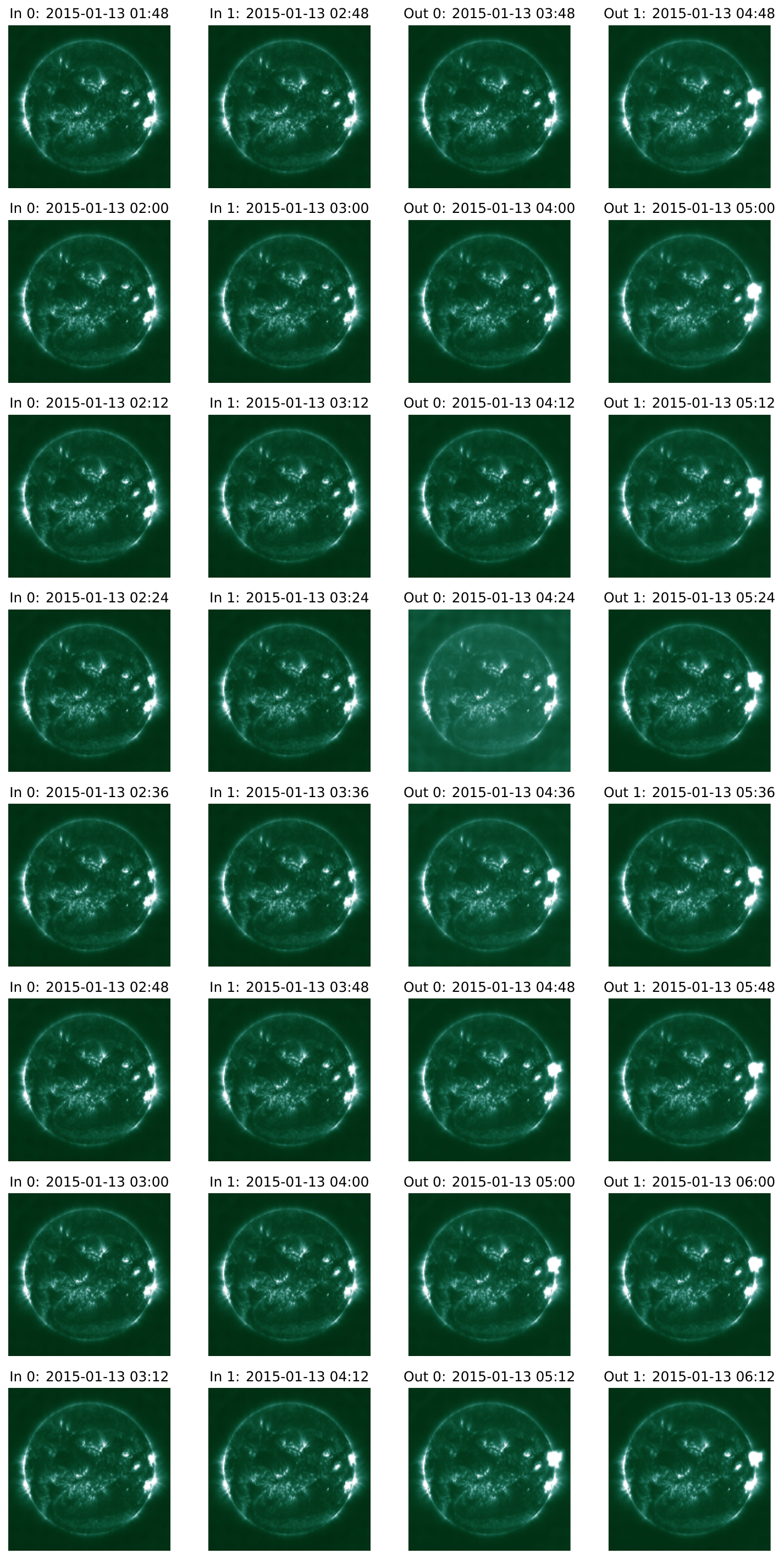} 
\caption{Surya inputs and outputs with different initializations for January 13, 2015.}
\label{fig:solar_flare_visual_overview_20150113}
\end{figure}

Figures \ref{fig:solar_flare_visual_overview_20150113} and \ref{fig:solar_flare_visual_overview_20140107} as well as \ref{fig:solar_flare_visual_20150113}, \ref{fig:solar_flare_model_run_2_20140107} and \ref{fig:solar_flare_model_run_4_20140107} answer this to the positive. They show model inputs, outputs, and ground truth at different initialization times on January 7 2014 and January 13 2015. While these are not exactly in our testing period of section \ref{sec:train_test_split} as they lie in the ``temporal buffer'', they are far enough from actual training data such that leakage and contamination is not a concern.

\begin{figure}[hp]
\centering
\includegraphics[width=1\textwidth]{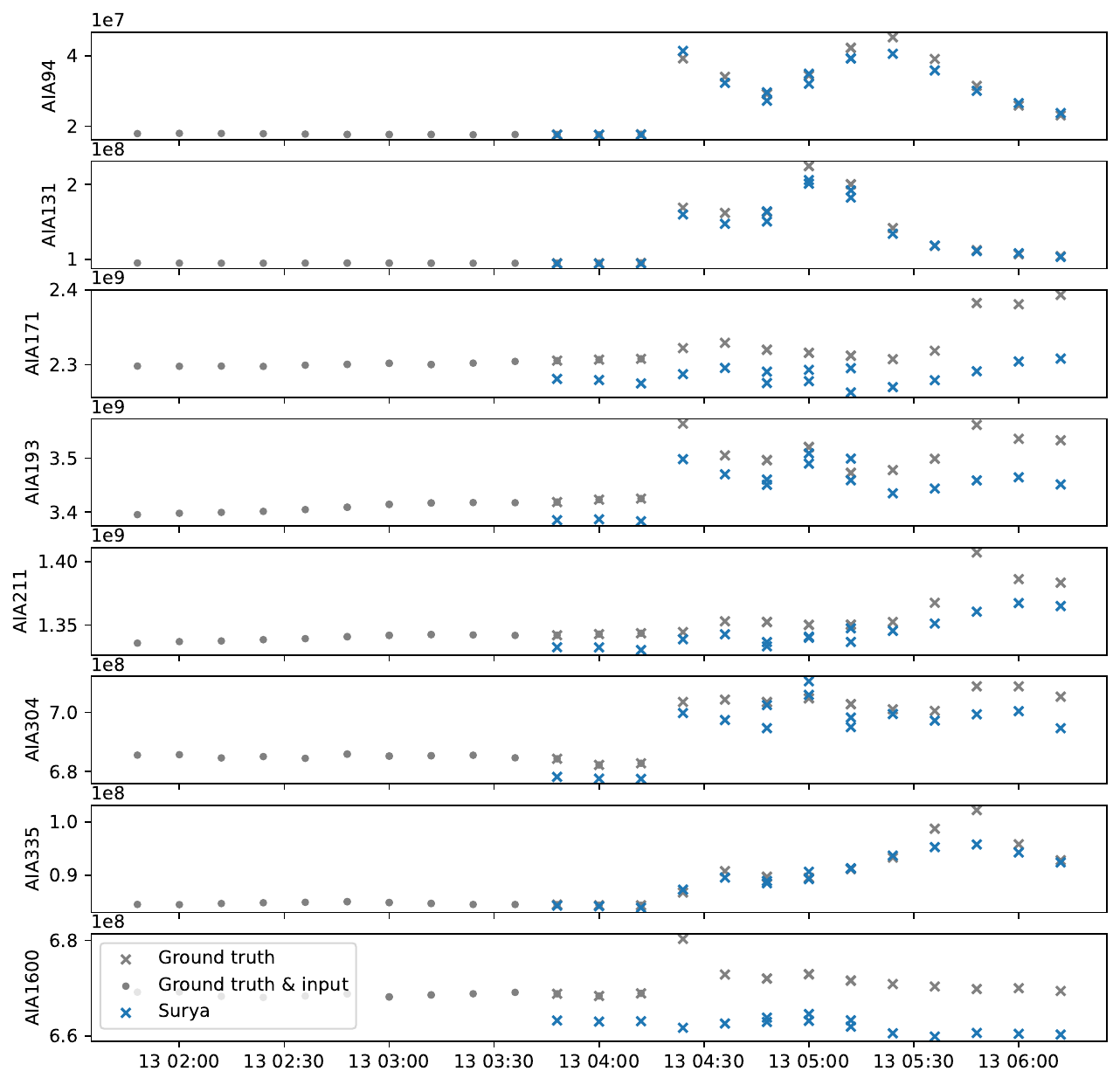} 
\caption{Integrated emissions for a solar flare event on January 13, 2015. That is, the plot shows the per-channel sum over all pixels in model inputs and outputs. The model was initialized for multiple lead times on January 13, 2015, and ran up to 2 hours (2 steps) into the future. The time series is the join of all those runs (which is why certain times show multiple outputs). This matches the visual output of figures \ref{fig:solar_flare_visual_20150113} and \ref{fig:solar_flare_visual_overview_20150113}.}
\label{fig:solar_flare_eve_emissions_20150113}
\end{figure}

The above figures show a strong visual feature evolving in channels 94 \AA and 131 \AA. Yet that alone does not allow us to characterize these as ``flare events''. To do that, we plot the integrated emissions for the January 13 2015 case in figure \ref{fig:solar_flare_eve_emissions_20150113}. Note that the time series here is a composite of multiple runs of Surya. Each was initialized with two timestamps 60 minutes apart (as usual) and run two steps (2 hours) into the future. %The most noteworthy point is that 
Notably, for all channels except 1600 \AA, the integrated output of Surya tracks the ground truth closely. In particular, close to the rapid changes around 04:24 UTC. We consider the results discussed in this section as \emph{visual} prediction of solar flare events. This is in contrast to the conventional approach discussed via fine-tuning in section \ref{sec:downstream_eval}.

\subsubsection{Blurring of sharp features and denoising of EUV channels}

\begin{figure}[htb]
\centering
\includegraphics[width=1\textwidth]{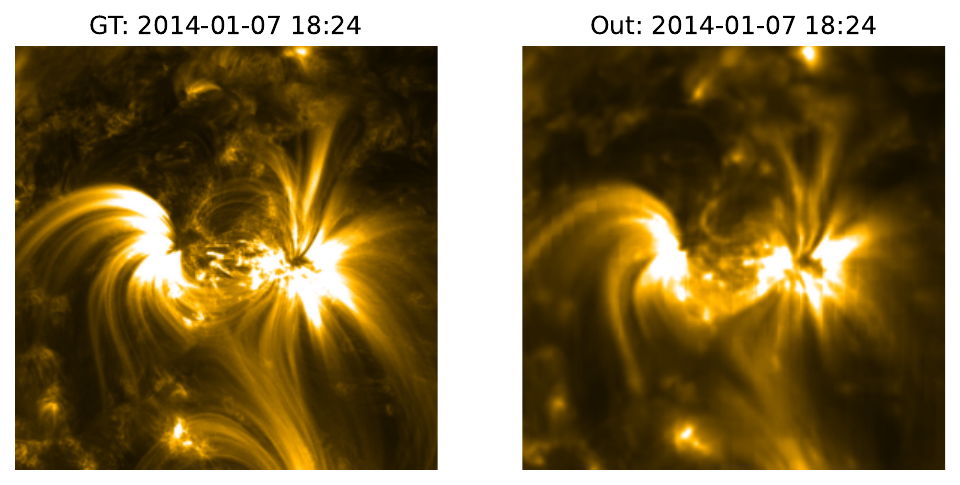} 
\caption{Ground truth data (left) and corresponding outputs from Surya for one hour ahead forecasting. 171\AA. The images shows a local crop of 600 by 600 pixel. }
\label{fig:detail_plot_20140107_2_aia171}
\end{figure}

\begin{figure}[htb]
\centering
\includegraphics[width=1\textwidth]{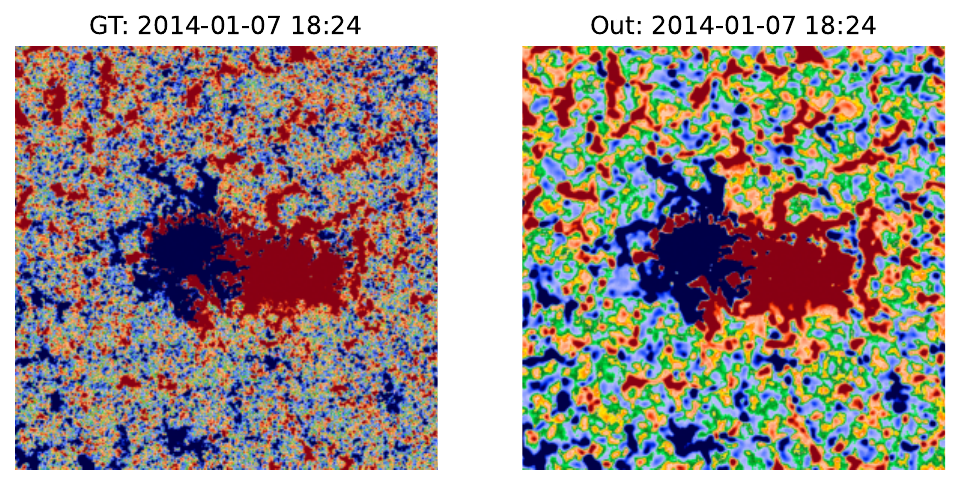} 
\caption{Ground truth data (left) and corresponding outputs from Surya for one hour ahead forecasting. HMI\_m. These images show a local crop of 600 by 600 pixels.}
\label{fig:detail_plot_20140107_2_hmi_m}
\end{figure}

Surya is trained as a fully deterministic model with an MSE objective. A well-known characteristic of such models is that they tend to blur sharp features. Figures \ref{fig:detail_plot_20140107_2_aia171} and \ref{fig:detail_plot_20140107_2_hmi_m} show local crops of Surya's outputs AIA 171 \AA as well as HMI\_m compared to ground truth. Given the aforementioned choices regarding Surya's architecture and pretraining objective, we observe indeed a loss of the finest details for HMI data as well as some blurring in the AIA bands. In theory, one might be able to address this by transitioning to a probabilistic model. This could be done via diffusion techniques, or via noise injection and a suitable loss function as in \cite{lang2024aifs}. We experimented with the latter, injecting noise into the model during the long-short attention blocks using adaptive layer norm and training on a CRPS objective. Yet we found that while it did yield improved detail, in particular in the HMI channels, it lead to occasional token-level artifacts.

While Surya blurs the sharp features in the high frequency, high signal-to-noise AIA 171 \AA, it has a positive denoising effect on low-signal to noise channels like AIA 94 \AA and AIA \AA 335.   These channels, observing the hottest and most energetic plasma in the solar atmosphere, tend to suffer from a low photon count and high amounts of noise.   Surya helps make this data more useful for analyzing small-scale features.

\subsection{Downstream Evaluation}
\label{sec:downstream_eval}

\subsubsection{Fine-tuning architecture and fine-tuning protocol}

\paragraph{Fine-tuning architecture} As discussed in section \ref{sec:forecasting_as_pretext}, Surya was pretrained with a forecasting pretext task. This was motivated by the dynamical properties of the solar surface and atmosphere
%solar system 
as well as the fact that some of our downstream tasks have a clear forecasting flavor. 
On the other hand, this can lead to a challenge when fine-tuning the model with frozen weights. The issue is that the representation learned by pretraining -- the activations of the last long-short attention layer -- can be assumed to be inherently \emph{local}. After all, in pretraining, we apply a linear layer to these activations to regress on the image seen by SDO at this specific location in the future. This is in contrast 
%difference 
to a masked reconstruction approach. Here, the encoder learns a representation from which the decoder can reconstruct the entire image. This implies that each token learns a representation which -- collectively with the other tokens -- can be used for global reconstruction.\footnote{The reader might complain that the representation learned by a masked autoencoder is still local. And indeed, if one plots the activations returned by the encoder, they still contain the data at that location. So one should take the above with a grain of salt. Still, the main point is that a model that regresses on future state will inherently be forced to contain local information in its ultimate transformer layers.} With this in mind, we implement multiple fine-tuning architectures for Surya.

We consider problems covering global classification (flare forecasting) and regression (solar wind and EUV spectral forecast). Here, we enable global average pooling, global max pooling, attention pooling, transformer pooling, and finally the use of a global class token. Let us briefly discuss each of these:

Global average and max pooling are straightforward: One aggregates the activations of the last transformer block and applies one or several linear layers. It is here where the comments from the preceding paragraph most apply: For a frozen model trained on a forecasting task, global max pooling can be assumed to return the most prominent (brightest) pixels in the output. And indeed, if one uses global max or global average pooling with frozen weights for solar flare forecasting, one obtains reasonable performance quickly; yet said performance quickly reaches a maximum that can be surpassed by approaches that really consider patterns rather than the maximal local activation values.

Attention pooling applies another attention layer before summing activations. Transformer pooling introduces an additional attention block with a dedicated class token. For downstream tasks that concern rare events -- see solar flare forecasting -- we find that attention and transformer pooling introduce too many parameters and overfit heavily.

Finally, we consider the global class token. Here, we simply introduce a non-local token after the spectral gating layers. This token is initialized with learnable weights. In contrast to Swin-transformers, an advantage of the long-short attention codebase is that the introduction of this class token is relatively straightforward.

\subparagraph{Parameter-efficient fine-tuning (LoRA)}

Note that all of the above can be combined with LoRA fine-tuning of the model. And indeed, this is how we obtain our strongest fine-tuning results. In other words, to adapt \textit{Surya} efficiently, selected linear maps (e.g., attention projections and MLP layers) are augmented with a low-rank residual while the pretrained weights remain frozen. For any targeted weight $W_0\in\mathbb{R}^{d_{\mathrm{out}}\times d_{\mathrm{in}}}$, LoRA parameterizes
\begin{equation}
W \;=\; W_0 \;+\; \frac{\alpha}{r}\,B A,
\qquad
A\in\mathbb{R}^{r\times d_{\mathrm{in}}},\;
B\in\mathbb{R}^{d_{\mathrm{out}}\times r},
\label{eq:lora}
\end{equation}
with rank $r$ and scaling $\alpha$. During fine-tuning, only the adapter parameters $(A,B)$ and any task-specific head(s) are updated; $W_0$ is fixed. Let $\mathcal{S}$ index the set of adapted layers. The resulting optimization problem is
\begin{equation}
\min_{\ \theta_{\mathrm{head}},\,\{A_\ell,B_\ell\}_{\ell\in\mathcal{S}}}
\ \ \mathcal{L}_{\mathrm{total}}\!\big(\theta_0,\theta_{\mathrm{head}},\{A_\ell,B_\ell\}\big)
\;+\;
\lambda_{\mathrm{lora}}\sum_{\ell\in\mathcal{S}}
\!\big(\|A_\ell\|_F^2+\|B_\ell\|_F^2\big),
\label{eq:lora-obj}
\end{equation}
where $\lambda_{\mathrm{lora}}$ regularizes the low-rank updates. This reparameterization can be interpreted as learning a task-specific, low-dimensional perturbation in the local tangent space of $W_0$, preserving the inductive biases of the pretrained model while controlling adaptation capacity via $r$ and $\alpha$.

\subsubsection{Active region Segmentation}
\begin{table}[H]
\centering
\caption{AR Segmentation results comparing baseline models with Surya}
\begin{tabular}{lccc}
\toprule
\textbf{Model} & \textbf{Params} & \textbf{IoU} & \textbf{Dice Coeff} \\
\midrule
Unet   & 9.2 M & 0.688 & 0.801 \\
Surya  & \textbf{4.1 M} & \textbf{0.768} & \textbf{0.853} \\
\bottomrule
\end{tabular}
\end{table}

\begin{figure}
    \centering
    \includegraphics[width=\linewidth]{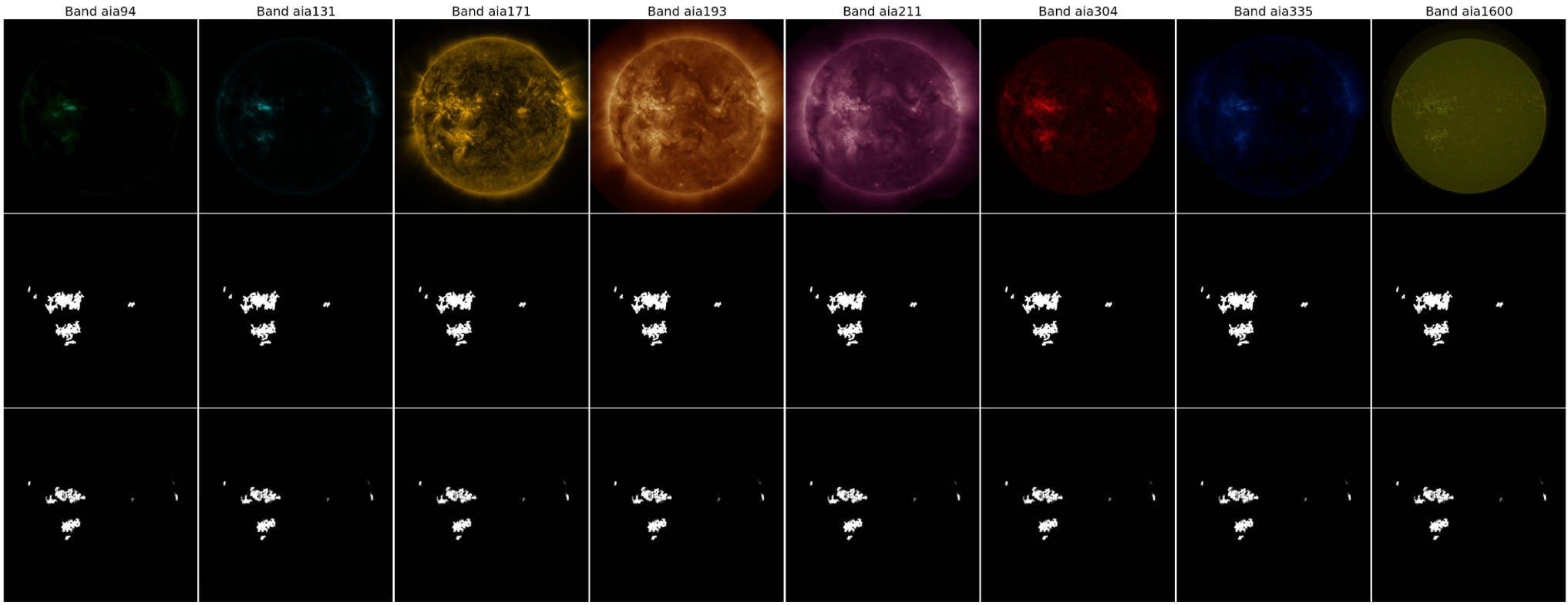}
    \caption{Active region Segmentation Results. Top Row: SDO data on Date: 2014-02-01, Time: 08:12, Middle Row: Model Output segmentation mask, Bottom Row: Ground Truth}
    \label{fig:arsegmentation results}
\end{figure}

Solar Active Regions (ARs) are magnetically complex structures associated with flares and CMEs. A key feature within ARs is the Polarity Inversion Line (PIL), the boundary separating opposite magnetic polarities, whose strong and sheared forms are robust precursors of eruptions \cite{Ji2023}. Accurately segmenting ARs containing PILs is thus critical for space weather forecasting and advancing our understanding of solar magnetic complexity.

Traditional AR/PIL detection pipelines based on thresholding and morphology are interpretable but brittle—sensitive to noise, parameter choices, and unable to capture the thin, filamentary structures of PILs. This motivates a deep learning–based segmentation framework that learns robust, multi-scale representations directly from solar data.

We construct the ARPIL dataset using full-disk SDO/HMI line-of-sight magnetograms ($4096 \times 4096$), adapting the method in \cite{Cai2020}. Positive and negative polarity maps are generated using $\pm 50$ G thresholds, filtered to remove regions smaller than 100 pixels, dilated with a 10-pixel kernel, and intersected to extract PILs. Only ARs with PILs are retained, yielding 119,454 binary masks spanning January 2011–December 2024.

As baselines, we compare a standard UNet with our finetuned Surya foundation model using LoRA adaptation. With just 4.1M trainable parameters, Surya achieves higher segmentation quality (IoU 0.768, Dice 0.853) than UNet (IoU 0.688, Dice 0.801).

\subsubsection{Solar Flare Forecasting}
As part of our downstream application evaluation, solar flare prediction is posed as a binary classification problem, where the objective is to determine whether a significant flare (M- or X-class) will occur within the next 24 hours following a time-point observation at $t$, similar to \cite{Pandey2023}. The prediction window is defined as $[t, t+24h)$ hours. The instance is labeled positive (flaring) if the peak X-ray flux of the strongest flare in $[t, t+24h)$ exceeds $\theta_{\text{max}} = 10^{-5}$ W/m\textsuperscript{2}.

% In case of Maximum Flare Intensity, the input is labeled positive if the strongest flare in $[t, t+2)$ exceeds $\theta_{\text{max}} = 10^{-4}$ W/m\textsuperscript{2} and in case of Cumulative Flare Intensity, the input is labeled positive if the sum of all flare intensities in $[t, t+2)$ surpasses $\theta_{\text{sum}} = 10$.

Given solar observations $\mathbf{x}_t \in \mathbb{R}^{C \times H \times W}$ (or $\mathbb{R}^{T \times C \times H \times W}$ for temporal sequences), the classifier predicts:

\begin{equation}
f(\mathbf{x}_t) \rightarrow y_t \in \{0,1\},
\end{equation}

with 
\[
y_t =
\begin{cases}
1, & \text{if a strong flare occurs in } [t, t+24h) \\
0, & \text{otherwise.}
\end{cases}
\]

We considered True Skill Statistic (TSS), Heidke Skill Score (HSS) and F1 score as evaluation metrics. TSS measures the ability to distinguish between flare and non-flare events. Ranges from -1 (inverse prediction) to +1 (perfect prediction), with 0 indicating no skill, and is defined as
\begin{equation}
\mathrm{TSS} \;=\; \frac{TP}{TP + FN} \;-\; \frac{FP}{FP + TN}.
\end{equation}

HSS evaluates performance relative to random chance, considering both hits and false alarms. Ranges from -1 
%\(-\infty\) 
to 1, and is defined as:
\begin{equation}
\mathrm{HSS} \;=\; 
\frac{2 \big( TP \cdot TN - FP \cdot FN \big)}
     {(TP + FN)(FN + TN) + (TP + FP)(FP + TN)} .
\end{equation}

F1 Score is the harmonic mean of precision and recall of positive class (i.e., flaring), balancing both false positives and false negatives, defined as :
\begin{equation}
\mathrm{F1} \;=\; \frac{2TP}{2TP + FP + FN}.
\end{equation}

The results of our evaluation are shown in Table~\ref{tab:flarepredresults} and we observe that Surya outperforms the baseline models across all evaluation metrics. Notably, Surya achieves superior results despite using fewer trainable parameters. Its advantage is evident in the balanced forecast skill metric (HSS), highlighting its robustness in handling class imbalance in solar flare prediction. It should also be noted that these models serve as proof-of-concept studies and are not optimized for end-to-end or operational forecasting use.

\begin{table}[H]
\centering
\caption{The results of flare forecasting evaluation comparing baseline models used in literature \cite{Pandey2023, Pandey2024} with finetuned Surya-based ones}
\label{tab:flarepredresults}
\begin{tabular}{lccc}
\toprule
\textbf{Model} & \textbf{TSS} & \textbf{HSS} & \textbf{F1} \\
\midrule
AlexNet        & 0.358  & 0.398  & 0.454 \\
ResNet50       & 0.018  & 0.028  & 0.055 \\
Surya  & \textbf{0.436} & \textbf{0.522} & \textbf{0.561} \\
\bottomrule
\end{tabular}
\end{table}

\subsubsection{Solar EUV spectra modeling}

Accurately modeling solar Extreme Ultraviolet (EUV) irradiance is essential for advancing space weather prediction, as it directly influences satellite functionality, communication infrastructures, and navigation systems. The challenge lies in modeling irradiance over 1343 spectral bands~\citep{eve_woods} ranging from 5 nm to 35 nm, which encode intricate spatial and temporal dynamics of multi-million degree solar plasma captured through observations.

Given multi-channel solar imagery $\mathbf{x}_t \in \mathbb{R}^{C \times H \times W}$ at time $t$, the objective is to estimate a continuous vector of EUV irradiance measurements $y_t \in \mathbb{R}^{1343}$:

\begin{equation}
    y_t = f(\mathbf{x}_t).
\end{equation}

\begin{table}[H]
\centering
\caption{EUV spectra modeling results comparing baseline models with Surya}
\label{tab:everesults}
\begin{tabular}{lccc}
\toprule
\textbf{Model} & \textbf{MSE} & \textbf{MAE} & \textbf{MAPE} \\
\midrule
Alexnet & 0.0001311409 & 0.0061858603 & 1.6834715604 \\
ResNet50 & 0.0030414662 &  0.0529725812 & 8.8635520935 \\
Surya & \textbf{0.0001260741} & \textbf{0.0045114677} & \textbf{1.4792510271} \\
\bottomrule
\end{tabular}
\end{table}

Table \ref{tab:everesults} shows the performance of EUV spectra modeling of surya compared to base models.
We find that the EUV irradiance prediction across the full spectrum from Surya model outperforms our baseline AlexNet and ResNet50 models. We note that the current operational model, the Flare Irradiance Spectral Model \citep[FISM,][]{Chamberlin2020}, attains a MAPE of 1.5\% and MSE of 0.00031 on the same dataset where Surya is tested, which is 2.46$\times$ worse than the Surya prediction. Although the MAPE of Surya is comparable to FISM, it is important to note that FISM relies on multiple instruments: the EUV Variability Experiment \citep{Woods2010}, the Solar Radiation and Climate Experiment \citep{Mcclintock2005}, and the XUV Photometer System \citep{Woods2005}, as well as three distinct proxies (the Penticton F10.7 cm radio flux, the Mg II core-to-wing index, and H I Lyman-$\alpha$ emission at 121.6 nm) spanning radio to X-ray wavelengths, while Surya relies solely on SDO/AIA and HMI data. The lower MSE achieved by Surya indicates enhanced accuracy in capturing extreme events, such as large flares. Furthermore, we also note that the Surya model outperforms the deep learning model by \cite{Szenicer19}, who obtain a median MAPE of 1.6\% and max MAPE of 4.6\%. We also note that while \cite{Szenicer19} predict the intensities in specific spectral lines, Surya generates predictions across all spectral bins of EVE, providing a full spectral readout. Hence, the Surya predictions are more comprehensive and better compared to the deep learning model by \cite{Szenicer19} reported in literature. 

\subsubsection{Solar Wind Forecasting}

Solar wind forecasting aims to predict the solar wind speed at a given spatial point (typically at the L1 point in the Sun-Earth system), specifically within a 4-day prediction window following an observation time $t$. Precise forecasting of solar wind speeds is fundamental for mitigating the adverse effects of space weather on satellite communication systems, navigation systems, and electrical grids on Earth.

This dataset comprises scalar measurements of solar wind speeds near Earth, recorded hourly from 2010-01-01 through 2023-12-31, resulting in a temporally rich dataset with substantial coverage of solar cycles~\citep{ace}. The solar wind speed values exhibit significant variability, ranging from $2.4 \times 10^{2}$ km/s to $8.8 \times 10^{2}$ km/s.

Given solar observation data (such as AIA and HMI multi-channel solar imaging data) represented by $\mathbf{x}_t$ at observation time $t$, the task is to predict the scalar solar wind speed at time $t + \Delta t$, where $\Delta t = 4$ days:

$$
y_{t+\Delta t} = f(\mathbf{x}_t),
$$

where $\mathbf{x}_t \in \mathbb{R}^{C \times H \times W}$ represents the multi-channel, high-resolution input imagery data at time $t$, and $y_{t+\Delta t} \in \mathbb{R}$ represents the predicted scalar solar wind speed. We select a forecast horizon of $4$ days corresponding to the typical travel time of solar wind plasma~\citep{CorHole_Rotter, Upendran20}.
The table below mentions various metrics by AlexNet, trained on 18k steps, ResNet50 for 20k steps, and Surya for 10k steps. Surya outperformed the baseline, achieving  the best results in the least number of steps

\begin{table}[H]
\centering
\caption{Solar Wind Prediction results comparing baseline models with Surya}
\begin{tabular}{lccc}
\toprule
\textbf{Model} & \textbf{RMSE }(\rm{km} $\mathrm{s}^{-1}$) & \textbf{MAE}(\rm{km} $\mathrm{s}^{-1}$) & \textbf{Validation MSE} ($\mathrm{km}^2 \mathrm{s}^{-2}$) \\
\midrule

AlexNet   & 118.6  & 95.7 km  &13839.49  \\
ResNet50  & 93.76 & 74.65 & 8547.924 \\
Surya & \textbf{75.92} & \textbf{58.06} & \textbf{5698.62} \\
\bottomrule
\end{tabular}
\end{table}

We may perform a qualitative comparison of the solar wind forecast from Surya with the performance of a few models reported in the literature. Each of these models considers data from the same instrument, while they are evaluated over a range of time periods. These results are summarized in Table~\ref {tab:swcomparison}.

\begin{table}[H]
\centering
\caption{Solar wind speed prediction results comparing baseline models with Surya}
\label{tab:swcomparison}
\begin{tabular}{lccc}
\toprule
\textbf{Model} & \textbf{RMSE }(\rm{km} $\mathrm{s}^{-1}$)\\
\midrule
Surya & \textbf{75.92} \\
WindNet 193 (1,4) \cite{Upendran20} & 84.33 $\pm$ 2.31 \\
WindNet 211 (1,4) \cite{Upendran20} & 85.94 $\pm$ 4.67 \\
27-day Persistence \cite{anh_2025_sw} & 89\\
SDO + solar wind speed model \cite{anh_2025_sw} & 68.6 \\
CH features + solar wind speed model: \cite{collins_2025_sw} & 61 -- 71 \\
MHD selected rotations \cite{swasti_mhd} & 47 -- 85 \\
ESWF using CH features: \cite{reiss2016verification}  & 108.2 \\
WSA model using photospheric B maps: \cite{reiss2016verification} & 99.5 \\
\bottomrule
\end{tabular}
\end{table}

From Table~\ref{tab:swcomparison}, we find that the Surya solar wind forecast outperforms the models that operate on purely SDO/AIA data or photospheric B maps, and the standard 27-day persistence baseline. We also note that the MHD model selected is evaluated on selected Carrington rotations, with the parameters fine-tuned for each rotation. This would correspond to overfitting a model on the testing set samples, meaning the selected parameters are not generalized across the dataset. Hence, the Surya model both generalizes across the full solar cycle and beats several dedicated solar wind speed forecasting models from the literature. The models that show better performance than Surya also consider the solar wind speed measurement at a previous time step, and hence contain a measure of memory of the state of the wind speed, rendering a comparison with Surya unfair. However, we note that even in this case, Surya's performance is close to said models, bringing it one step closer to being an effective operational model.

\section{Discussion and conclusions}

In this work, we presented Surya, a 366M-parameter foundation model for heliophysics, trained at the native 4096×4096 resolution of SDO’s AIA and HMI instruments with a standardized 12-minute cadence. By pretraining on the task of forecasting, Surya learns general-purpose solar representations that capture both the fine-scale variability of magnetic fields and the large-scale dynamics of the solar atmosphere. This pretraining strategy enables the model to perform zero-shot forecasting of solar activity, including the visual evolution of flare events, while also providing transferable representations that can be adapted efficiently to various downstream applications. Surya thus represents a shift from narrowly focused, task-specific models to a more versatile and scalable approach for heliophysics. This positions Surya as a backbone for future digital twin initiatives of the Sun-Earth system, aligned with ESA’s Digital Twin Earth and NASA’s digital twin efforts.

A key result is Surya’s capability to forecast solar dynamics without additional training. For example, in the case of the January 2015 flare, Surya’s predicted integrated EUV emissions closely tracked observations, demonstrating sensitivity to the rapid changes associated with flare onset. Quantitatively, autoregressive rollout tuning improved long-range forecasting skill by up to 17.8\% at 12 hours lead time compared to one-step pretraining. These results suggest that the model is not simply memorizing past patterns, but rather developing representations that are, to some extent, physics-aware. It also significantly outperforms persistence (MAE $\approx$ 0.59) and learned-flow baselines (MAE $\approx$ 0.34) in one-hour forecasts.

For finetuning on downstream tasks, we adapted parameter-efficient fine-tuning with LoRA. For active region segmentation, Surya achieved an IoU of 0.768 and a Dice coefficient of 0.853, outperforming a U-Net baseline (IoU 0.688, Dice 0.801). In solar flare forecasting, Surya obtained TSS of 0.436, HSS of 0.522, and F1 of 0.561, substantially improving over AlexNet (TSS = 0.358) and ResNet50 (TSS = 0.018). For EUV irradiance modeling, it reduced error with an MAE of 0.0043 compared to 0.0065 for AlexNet and 0.0256 for ResNet50. Finally, in solar wind speed forecasting, Surya achieved RMSE of 75.92 \rm{km} $\mathrm{s}^{-1}$ and MAE of 58.06 \rm{km} $\mathrm{s}^{-1}$, outperforming both AlexNet (RMSE = 118.6 \rm{km} $\mathrm{s}^{-1}$) and ResNet50 (RMSE = 93.76 \rm{km} $\mathrm{s}^{-1}$). We further note that Surya beats models from literature that use SDO/AIA as the sole input, while also outperforming the 27-day persistence model. The model performance is close to the performance of models that consume prior solar wind speed measurements and MHD models, which show the power of generalization demonstrated by the foundation model. However, we do note caution against over-interpretation of this comparison, since the validation dataset is typically not precisely the same timestamps across all the models.

We also observed some limitations that can be improved in future work. As a deterministic model trained with an MSE objective, Surya exhibits blurring of sharp features in magnetograms and flare imagery, a common outcome in regression-based generative models. Probabilistic approaches, such as diffusion forecasting or training with alternative loss functions like CRPS, could mitigate this issue by producing sharper, more physically realistic predictions. Similarly, while forecasting proved effective as a pretraining task, other self-supervised strategies such as masked reconstruction or band-to-band translation may further enrich the learned representations. The main bottleneck was not model-scaling but data throughput during rollout training, suggesting that future large-scale heliophysics FMs will require investments in data pipelines as much as compute.

In conclusion, Surya represents the first foundation model for heliophysics trained at the full resolution of SDO data, establishing a unifying framework that combines forecasting skill with transferable representations for a range of scientific and operational tasks. Its ability to generalize across segmentation, classification, regression, and forecasting problems illustrates the potential of foundation models to accelerate both discovery and operational space weather prediction. Looking forward, we can incorporate multimodal, multi-mission datasets and adopt probabilistic approaches for better and improved foundation models to support next-generation heliophysics and digital twin initiatives.

\section*{Code and Data Availability}
By releasing Surya, its preprocessing pipeline, and downstream evaluation code openly, we align with NASA’s Year of Open Science and the NAIRR pilot, ensuring that the community can build upon this foundation. The model and datasets are publicly available on Huggingface: \url{https://huggingface.co/nasa-ibm-ai4science}. Our code for model and downstream tasks is publicly available at \url{https://github.com/NASA-IMPACT/Surya}

\subsubsection*{Acknowledgments}

We would like to thank Soumya Ranjan and WeiJi Leong from Development seed who contributed in the early stages of this project. We would also like to thank Shubha Ranjan from NASA Advanced Supercomputing (NAS) Division, and Mike Little from Goddard Spaceflight Center for their help and support. We would also like to thank David Hall for help and support with Nvidia computing resources.

The Authors acknowledge the National Artificial Intelligence Research Resource (NAIRR) Pilot and NVIDIA for providing support under grant no.~NAIRR240178. The authors would also like to thank NASA Advanced Supercomputing (NAS) Division for their compute support. Vishal Upendran would like to acknowledge NASA for support under award number 80NSSC25K7956.  NVP and TS have been supported, in part, by NASA R2O2R grant 80NSSC22K0270. DVH is grateful for the support provided by the NASA FINESST grant 80NSSC22K0058.

\bibliography{iclr2025_conference}
\bibliographystyle{iclr2025_conference}

\appendix

\section{Architecture}

\subsection{Ablation studies}
\label{sec:ablation_studies}

The main body of the text already discusses the baseline scores: persistence as well as the flow model trained to capture differential rotation. In addition, we performed a number of ablation studies to validate Surya's architecture choices. All ablation experiments were trained for 10,000 gradient descent steps on 16 GPUs. The results can be found in table \ref{table:Ablation studies}.

\begin{table}[htb]
    \centering
    \caption{Architecture ablations \& baselines. The table shows MSE loss for 1 hour ahead forecasting in model units of \eqref{eq:signum_log_scaling}. All baselines and ablations in this table -- including the Surya configuration -- were trained on 16 GPUs for 10,000 gradient descent steps.}
    \label{table:Ablation studies}
    \begin{tabular}{llrrr}
        \toprule
        \textbf{Type} & \textbf{Configuration} & \textbf{Parameters} & \textbf{Memory usage [MiB]} & \textbf{Loss (MSE)} \\
        \midrule
        Baseline & Persistence & N/A & N/A & $0.594044030$ \\
        & Learned flow & $642$ & $24976$ & 0.337624282 \\
        \midrule
        Ablation & Single timestamp & $361.93$ M & $55411$ & 0.228553504 \\
        & No spectral gating & $210.39$ M & $59823$ & $0.219618767$ \\
        & Perceiver & $351.52$ M & $52721$ & $0.234643415$ \\
        \midrule
        Surya &  & $366.19$ M & $56247$ & $0.219778508$ \\
        \bottomrule
    \end{tabular}
\end{table}

%The main body of the text already discusses the baseline scores: Persistence as well as the flow model trained to capture solar rotation. In addition, we performed a number of ablation studies to validate Surya's architecture choices. All ablation experiments were trained for 10,000 gradient descent steps on 16 GPUs. The results can be found in table \ref{table:Ablation studies}.

Let us start by discussing the spectral gating layers. In this ablation study (``No spectral gating''), we replace the two spectral gating layers with additional long-short attention layers. Within the context of the compute budget used here, this led to a virtually identical loss, yet at 6\% reduced GPU memory consumption. Interestingly, this happens although the conversion of long-short to spectral gating layers adds a huge number of parameters, each spectral block containing a weight matrix consisting of $84,541,440$ real parameters alone. Indeed, table \ref{table:Ablation studies} shows that a large number of Surya's parameters are in the two large weight matrices applied in Fourier space in those layers.

The next ablation we consider is training the model with one rather than two timestamps as input. Using two timestamps as input improves performance by 3.8\%. On the one hand, one clearly expects two timestamps to do better than one, as the model can infer a motion field. On the other one might be surprised that one timestamp as input still yields relatively strong performance. One reason for this might be that the tokenization in Surya is effectively a compression. Given two timestamps, 13 bands, and a patch size of 16 by 16, each patch comprises $2 \times 13 \times 16 \times 16 = 6,656$ pixels. Our embedding dimension, on the other hand, is $1,280$. So we are effectively compressing our data by a factor of 5.2 with a very simple linear layer. Using one timestamp as input rather than two reduces this compression ratio.

Given this compression ratio, one might expect that using a more complex tokenization procedure would help model performance. Indeed, one can consider the case of atmospheric physics where there is a similar situation: Model inputs comprise many different variables at different vertical levels. \cite{nguyen2024scaling} used an attention mechanism to aggregate variables. \cite{bodnar2024aurora} extended this to the use of a perceiver. The latter was also use in \cite{majid2024solaris}. With this in mind, we evaluated the use of a perceiver to aggregate input tokens and process outputs. As table \ref{table:Ablation studies}, we obtained considerably worse performance. Note however that the memory consumption of the perceiver was such that we had to use a 32 by 32 token size in this case. Which of course makes the problem worse that the perceiver was supposed to address.

Let us conclude with a few remarks about the flow model. To start, it is remarkable how well it performs given its miniscule parameter count. For AI forecasting and foundation models in atmospheric physics, it is very common not to model the target directly, but to model the difference of the target from some known quantity. In \cite{lam2022graphcast} the authors predicted the difference of the latest input timestamp $\mathbf{X}_t$ from the target $\mathbf{X}_{t+1}$. That is, the model was trained to predict the difference from persistence. Noting that in the presence of sharp features the model has to learn to exactly remove the sharp feature from $\mathbf{X}_t$ and add it at a new location in $\mathbf{X}_{t+1}$, \cite{lang2024aifs} improved on this by only considering the deviation from a smoothened version of $\mathbf{X}_t$. \cite{schmude2024prithvi} was interested in the case of zero lead time. Here, predicting a difference from the presence becomes meaningless and the authors chose to instead model the delta from historical climate. In either case, given the parameter efficiency of the flow model, it is tempting to first train a flow model and then a transformer to model the discrepancy between the flow model and the actual target. In our ablation studies, we found this to perform worse than simply regressing directly onto the target.

\section{Additional results}

\subsection{Impact of rollout tuning}

Table \ref{table:Rollout_tuning_details} shows the data corresponding to figure \ref{fig:rollout_eval}. Note that rollout tuning massively improves model performance at long lead times, we also see ever so slightly decreasing performance at the shortest ones. I.e.~the best performing model at 1 hour ahead lead time is actually Surya before phase 2 of pretraining.

\begin{table}[htb]
    \centering
    \caption{Impact of rollout tuning. The table shows validation loss (MSE) per lead time for lead times up to 12 hours ahead (rows). Columns show persistence and flow baselines as well as Surya at various stages of pretraining. The 5 hour ahead tuned version (rightmost column) corresponds to Surya as released. The best score in each row is marked in bold.}
    \label{table:Rollout_tuning_details}
    \begin{tabular}{rrrrrrrr}
        \toprule
        & \textbf{Persistence} & \textbf{Flow} & \textbf{1 hour (no rollout)} & \textbf{2 hours} & \textbf{3 hours} & \textbf{4 hours} & \textbf{5 hours} \\
        \midrule
        1 & $0.59868$ & $0.34204$ & $\mathbf{0.20133}$ & $0.20153$ & $0.20159$ & $0.20169$ & $0.20172$ \\
        2 & $0.70594$ & $0.40124$ & $0.22738$ & $0.22371$ & $0.22352$ & $0.22359$ & $\mathbf{0.22351}$ \\
        3 & $0.77862$ & $0.45497$ & $0.24889$ & $0.24182$ & $0.23928$ & $0.23902$ & $\mathbf{0.23889}$ \\
        4 & $0.83790$ & $0.50416$ & $0.26662$ & $0.25759$ & $0.25213$ & $0.25125$ & $\mathbf{0.25078}$ \\
        5 & $0.89444$ & $0.55291$ & $0.27976$ & $0.26957$ & $0.26295$ & $0.26141$ & $\mathbf{0.26064}$ \\
        6 & $0.94441$ & $0.59671$ & $0.29362$ & $0.28111$ & $0.27366$ & $0.27140$ & $\mathbf{0.26995}$ \\
        7 & $0.99986$ & $0.64647$ & $0.31217$ & $0.29536$ & $0.28557$ & $0.28113$ & $\mathbf{0.27854}$ \\
        8 & $1.04645$ & $0.68856$ & $0.32528$ & $0.30458$ & $0.29519$ & $0.29018$ & $\mathbf{0.28583}$ \\
        9 & $1.09013$ & $0.72833$ & $0.33984$ & $0.31209$ & $0.30361$ & $0.29777$ & $\mathbf{0.29216}$ \\
        10 & $1.12573$ & $0.76106$  & $0.35257$ & $0.31934$ & $0.31181$ & $0.30522$ & $\mathbf{0.29868}$ \\
        11 & $1.14940$ & $0.78280$  & $0.37037$ & $0.33154$ & $0.32443$ & $0.31638$ & $\mathbf{0.30789}$ \\
        12 & $1.16841$ & $0.79964$  & $0.38517$ & $0.34304$ & $0.33786$ & $0.32791$ & $\mathbf{0.31651}$ \\
        \bottomrule
    \end{tabular}
\end{table}

\subsection{Visual prediction of solar flares}
\label{sec:additional_visual_prediction_solar_flares}

\subsubsection{The 2024-01-07 event}

\begin{figure}[htbp]
\centering
\includegraphics[width=0.9\textwidth]{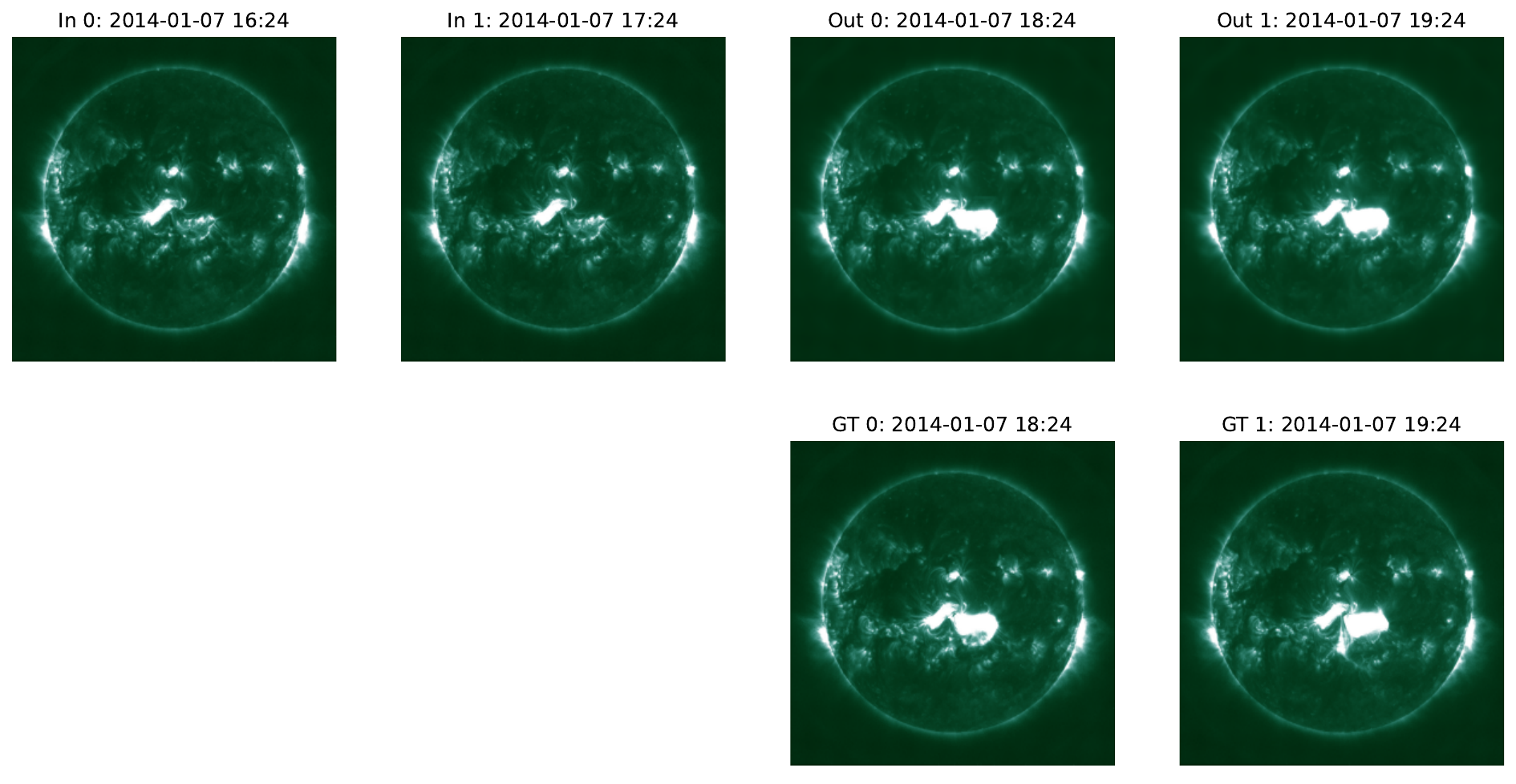} 
\caption{Surya inputs, outputs, and ground truth for January 7, 2014. AIA 94\AA. Top row, left two columns are model inputs (``In''). Top row, right two columns are model outputs (``Out''). Bottom row shows the corresponding ground truth (``GT''). The model is initialized slightly later than in \ref{fig:solar_flare_model_run_4_20140107}, so the flare is already visible in its first output frame 60 minutes ahead.}
\label{fig:solar_flare_model_run_2_20140107}
\end{figure}

\begin{figure}[htbp]
\centering
\includegraphics[width=0.9\textwidth]{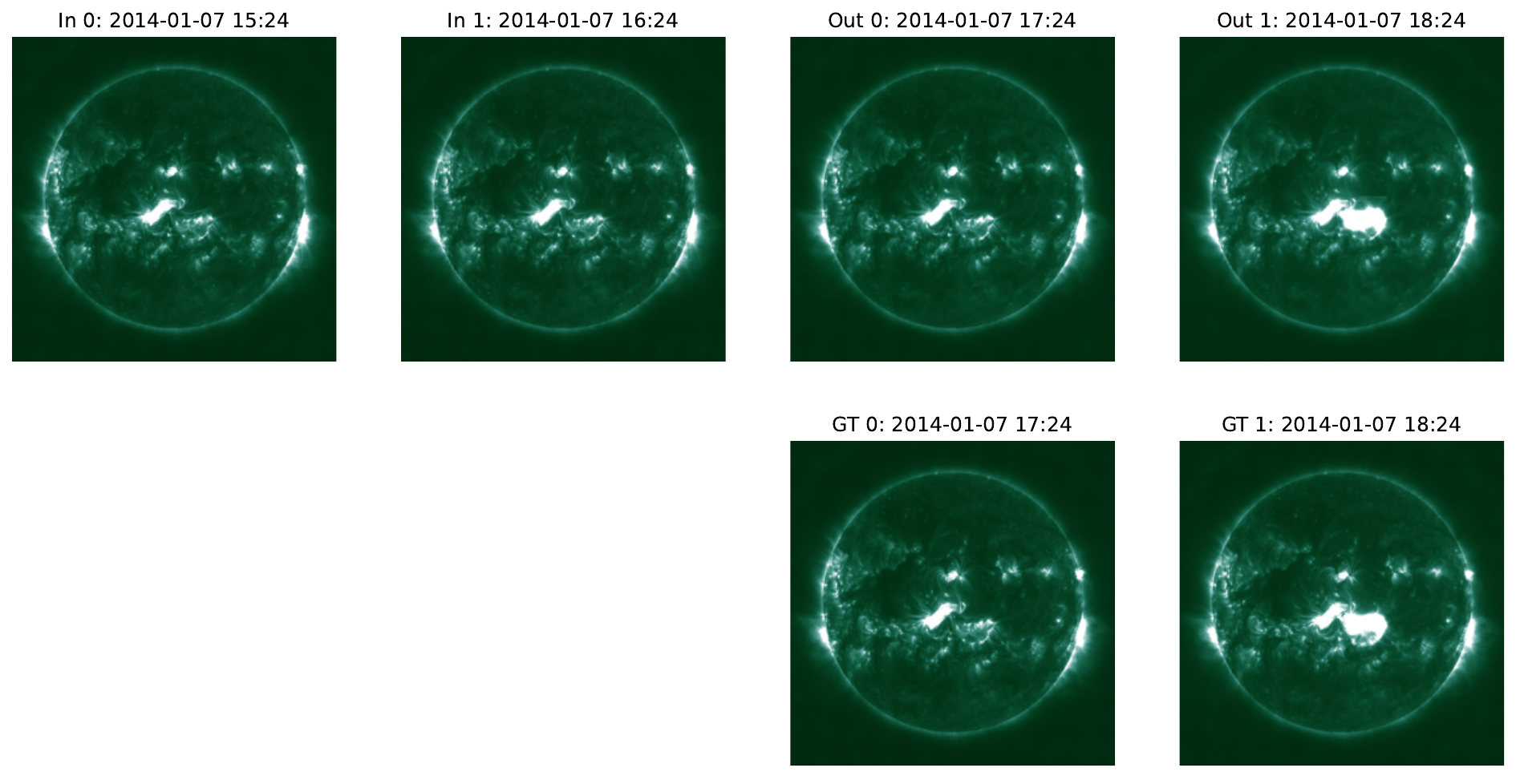} 
\caption{Surya inputs, outputs, and ground truth for January 7, 2014. AIA 94\AA. Top row, left two columns are model inputs (``In''). Top row, right two columns are model outputs (``Out''). Bottom row shows corresponding ground truth (``GT'').}
\label{fig:solar_flare_model_run_4_20140107}
\end{figure}

\begin{figure}[hp]
\centering
\includegraphics[height=0.9\textheight]{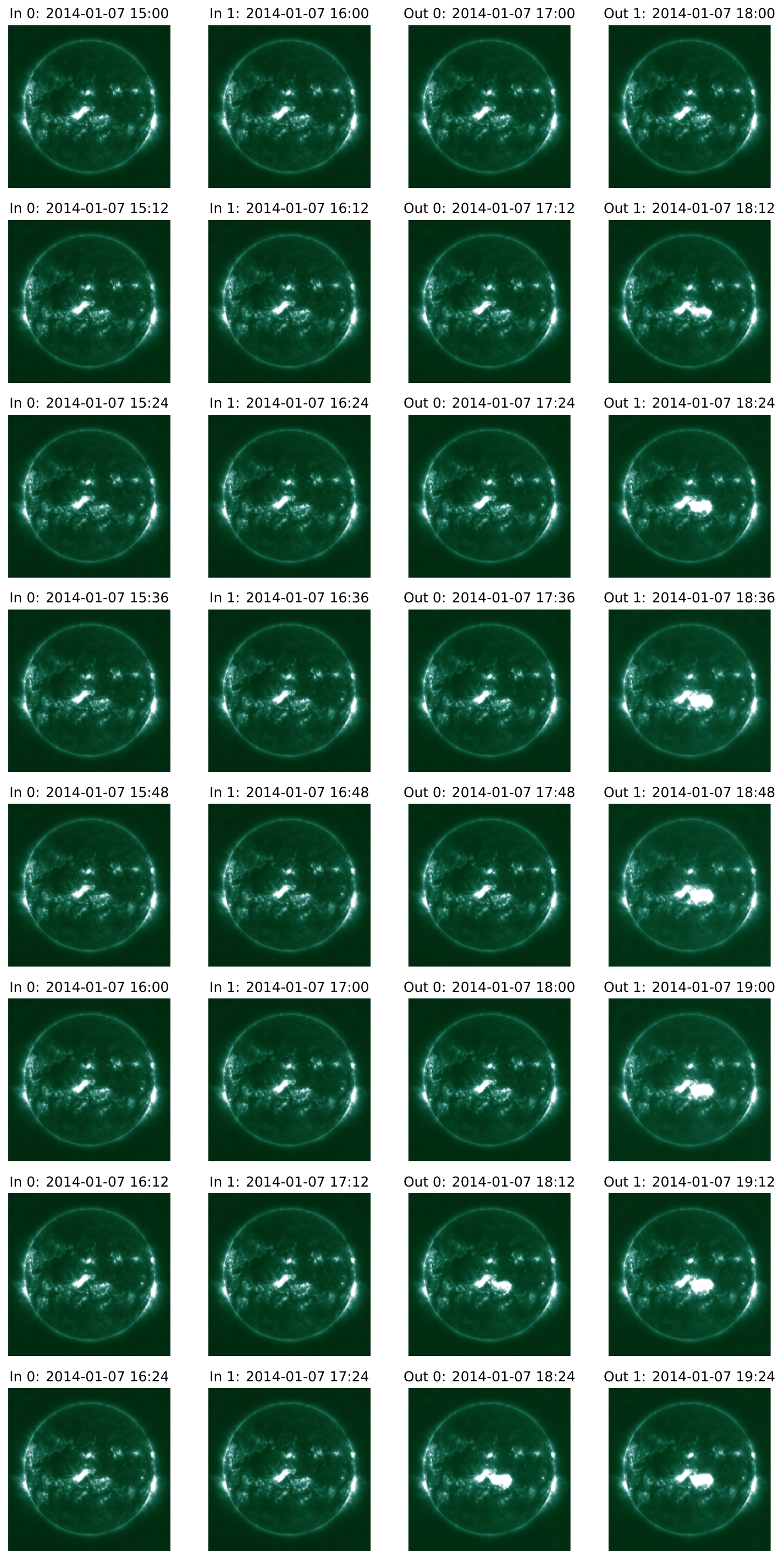} 
\caption{Surya inputs and outputs with different initializations for January 7, 2014. AIA 94\AA.}
\label{fig:solar_flare_visual_overview_20140107}
\end{figure}

Figures \ref{fig:solar_flare_model_run_2_20140107}, \ref{fig:solar_flare_model_run_4_20140107}, and \ref{fig:solar_flare_visual_overview_20140107} show model inputs and outputs as well as ground truth for a solar flare event on January 7, 2014. This is complementary to figures \ref{fig:solar_flare_visual_20150113} and \ref{fig:solar_flare_visual_overview_20150113}. Note that both cases show testing data. See section \ref{sec:train_test_split} for a discussion of the train/test split.

\end{document}